\documentclass[english,aps,prd,
groupedaddress,amssymb,
showpacs,nofootinbib]{revtex4}

\usepackage{times}
\usepackage{amsmath}
\usepackage{graphicx}

\newcommand{\beq}{\begin{eqnarray}}
\newcommand{\eeq}{\end{eqnarray}}

\makeatletter
\numberwithin{equation}{section}
\numberwithin{figure}{section}

\makeatother

\usepackage{babel}

\begin{document}
\setlength{\unitlength}{1mm}

\title{$(2+1)$-Dimensional  Yang-Mills Theory and Form Factor Perturbation Theory}

\author{Axel \surname{Cort\'es Cubero}}

\email{acortes_cubero@gc.cuny.edu}



\affiliation{ Baruch College, The 
City University of New York, 17 Lexington Avenue, 
New 
York, NY 10010, U.S.A. }

\affiliation{ The Graduate School and University Center, The City University of New York, 365 Fifth Avenue,
New York, NY 10016, U.S.A.}

\begin{abstract}
We study Yang Mills theory in 2+1 dimensions, as an array of coupled (1+1)-dimensional principal chiral sigma models. This can be understood as an anisotropic limit where one of the space-time dimensions is discrete and the others are continuous.
The $SU(N)\times SU(N)$ principal chiral sigma model in 1+1 dimensions is integrable, asymptotically free and has massive excitations. New exact form factors and correlation functions of the sigma model have recently been found by the author and P. Orland. In this paper, we use these new results to  calculate physical quantities in (2+1)-dimensional Yang-Mills theory, generalizing previous  $SU(2)$ results by Orland, which include the string tensions and the low-lying glueball spectrum. We also present a new approach to calculate two-point correlation functions of operators using the light glueball states. The anisotropy of the theory yields different  correlation functions for operators separated in the $x^1$ and $x^2$-directions.
\end{abstract}

\maketitle

\section{Introduction}

In this paper, we study an anisotropic version of (2+1)-dimensional Yang-Mills theory. The anisotropy is given by a longitudinal rescaling of coordinates of the  form $x^{0,1}\to \lambda x^{0,1}$, and $x^2\to x^2$. The gauge fields transform as $A_{0,1}\to(1/\lambda) A_{0,1}$, $A_2\to A_2$. The strength of the interactions is different in different directions. We explore the highly anisotropic regime, where $\lambda\to 0$.

We realize this rescaling by starting with the Kogut-Susskind Hamiltonian formulation of lattice gauge theory, with lattice spacing $a$. The rescaling of coordinates amounts to taking the continuum limit in the $x^0$ and $x^1$-directions, with the lattice spacing rescaling as $\lambda a$.

We view the anisotropic model as an array of two-dimensional field theories, coupled together to form a higher-dimensional theory. The strength of the coupling between these two-dimensional models depends on the rescaling parameter $\lambda$. The two-dimensional theory is the principal chiral sigma model (PCSM)\cite{twoplusone}. The PCSM is known to be integrable, and this property has been exploited to find  exact results \cite{integrablePCSM}, \cite{wiegmann}. The main goal of our program is to use exact results from the PCSM to calculate physical quantities in anisotropic QCD, finding corrections for small $\lambda$.

This anisotropic regime of Yang-Mills theory has been studied extensively by P. Orland. In Ref. \cite{twoplusone} it was established that the anisotropic theory is equivalent to an array of coupled PCSM's, and it was shown that the model confines quarks and has a mass gap.  In references \cite{horizontal}, and \cite{vertical} the string tensions for quark-antiquark pairs was found for the $SU(2)$ gauge group. In Ref. \cite{glueball} the low-lying glueball spectrum was found for $SU(2)$. In this paper we generalize all these results to all $N$. This is done using new form factors of the PCSM  that were found in Ref. \cite{multiparticle}. We also use the light glueball states to calculate  long-distance correlation functions of gauge-invariant operators.

The longitudinal rescaling of coordinates is inspired by a similar investigation in 3+1 dimensions by Verlinde and Verlinde \cite{verlinde}, in the context of heavy ion collisions. A similar anisotropic limit  was used by McLerran and Venugopalan in their derivation of the Color Glass Condensate picture \cite{MV}. An  anisotropic theory has been explored in Ref. \cite{miransky}, where the anisotropy is produced by an external magnetic field.

This approach is especially interesting for $(2+1)$-dimensional QCD, since there are two different coupling constants for the gluon field, but they are both 
small compared to the cutoff. This makes our approach fundamentally different from other analytic studies of $(2+1)$-dimensional QCD (which are generally at large dimensionless coupling) \cite{nair},\cite{greensite}. Recently, Karabali, Nair and Yelnikov \cite{KNY} have computed corrections to the results in \cite{nair}, in powers of the coupling constant. Their approach could eventually be used to study confinement at weak coupling.

Physical quantities in the anisotropic gauge theory can be evaluated in the context of form-factor perturbation theory \cite{DMS}, \cite{DM}, \cite{DGM}. The gauge theory with $\lambda= 0$ is integrable. The S-matrix, some form factors and correlation functions of the PCSM are known. We  do a perturbative expansion in powers of $\lambda$, rather than the Yang-Mills coupling constant. The perturbation theory  starts from an integrable, rather than a  free theory.

A very similar approach has been used by Konik and Adamov \cite{konikadamov}, and James and Konik \cite{konikjames} to examine the 3-dimensional Ising model as an array of 2-dimensional chains. Here they have successfully computed critical exponents and the entanglement entropy and spectra using an improved version of the truncated conformal spectrum approach.

In the next section, we present a discussion of the longitudinally-rescaled Yang-Mills theory. We show how the rescaled theory is equivalent to an array of integrable models. This equivalence is shown in the axial gauge on the Kogut-Susskind Hamiltonian.

In Section III, we compute the string tension for a static quark-antiquark pair separated in the $x^1$-direction only. In Section IV, we calculate the string tension of a quark-antiquark pair separated in the $x^2$-direction. These string tensions are different because the theory is anisotropic. In Section V we compute the mass spectrum of the lightest glueball states. These results generalize Orland's $SU(2)$ results to $SU(N)$.

In Section VI, we calculate the long-distance two-point correlation function of two gauge-invariant operators separated in the $x^1$-direction. This calculation is inspired by a similar calculation for the 2-dimensional Ising model in an external magnetic field, by Bhaseen and Tsvelik \cite{bhaseen}. 

In Section VII, we propose a method for calculating correlation functions in the $x^2$-direction. This is done by defining a transfer matrix that describes the evolution of the system along the $x^2$-direction. The partition function and correlation functions can be found, in principle, by diagonalizing the transfer matrix in the basis of physical states. We are only able to find an expression for the transfer matrix using the light glueball states from Section V. However, this matrix is very difficult to diagonalize. The problem is reduced to an integral eigenvalue equation, which we leave unsolved.

We present our conclusions in the last section. A short summary of the S-matrix and form factors of the PCSM is given in the appendix.

\section{Longitudinally Rescaled Yang-Mills Hamiltonian in the Axial Gauge}
\setcounter{equation}{0}
\renewcommand{\theequation}{2.\arabic{equation}}

In the Kogut-Susskind lattice Hamiltonian formulation \cite{kogut}, there are $SU(N)$-valued gauge fields $U(x)_j$, and electric-field operators $l(x)_{j}^b$ in the adjoint representation of $SU(N)$, at every space link $(x,j)$, for $j=1,2$ and $b=1,2,\dots, N^2-1$. These satisfy the commutation relations
\beq
\left[l(x)_{j}^b,l(y)_{k}^c\right]=i\delta_{x\,y}\delta_{j\,k} f^{dbc}l(x)_{j\,d},\,\,\,\,\,\,\,\left[l(x)_{ \,j}^b,\,U(y)_{k}\right]=-\delta_{x\,y}\delta_{j\,k}\,t^b U(x)_{j}.\nonumber
\eeq
The gauge fields in the $x^0$ direction are fixed by the temporal gauge condition $U_0=1$. The Hamiltonian is obtained by taking the continuum limit of the Wilson action in the time direction. The Kogut-Susskind Hamiltonian, inside a box of size $a^2 L_1\times L_2$, is
\beq
H=\sum_{x^1,x^2=-\frac{L_1}{2},\frac{L_2}{2}}^{\frac{L_1}{2},\frac{L_2}{2}}\sum_{j=1}^{2}\sum_{b=1}^{N^2-1}\frac{g_0^2}{2a}\left[l(x)_{j}^b\right]^2-\sum_{x^1,x^2=-\frac{L_1}{2},\frac{L_2}{2}}^{\frac{L_1}{2},\frac{L_2}{2}}\frac{1}{4g_0^2\,a}\,{\rm Tr}\,\left[U(x)_1U(x+\hat{1})_2U^\dag(x+a\hat{2})_1U^\dag(x)_2+C.C.\,\right].\label{kogutsusskind}
\eeq
where $L_1,\,L_2$ are even integers.

In temporal gauge, physical states, $\Psi$, are those which satisfy Gauss's law:
\beq
\sum_{j=1}^{2} [\mathcal{D}_j l_j(x)]_b \Psi=0,\label{latticegauss}
\eeq
where
\beq
[\mathcal{D}_j l_j(x)]_b=l_j(x)-\mathcal{R}_j(x-\hat{j}a)_b^{\,\,c}\,\,l_j(x-\hat{j}a)_c,\nonumber
\eeq
where $\mathcal{R}_j(x)_b^{\,\,c}t_c$ is the adjoint representation of the gauge field,
\beq
\mathcal{R}_j(x)_b^{\,\,c}t_c=U_j(x)t_b U^\dag_j(x).\nonumber
\eeq

We find the electric field component $l_1$ by solving Gauss's law (\ref{latticegauss}); and then  impose the axial gauge $U_{1}(x)={\mathbf 1}$, yielding
\beq
l_1(x^1,x^2)_b=\sum_{y^1=-\frac{L_1}{2}}^{x^1}[\mathcal{D}_2l_2(y^1,x^2)]_b.\label{latticeelectric}
\eeq
There is a global invariance left after the axial gauge fixing:
\beq
\sum_{x^1=-\frac{L_1}{2}}^{\frac{L_1}{2}}[\mathcal{D}_2l_2(x^1,x^2)]_b\Psi=0.\label{residualgausslattice}
\eeq

The lattice Hamiltonian in axial gauge is found by substituting the new nonlocal expression for the electric field (\ref{latticeelectric}) into (\ref{kogutsusskind}):
\beq
H&=&\sum_{x^1=-\frac{L_1}{2}}^{\frac{L_1}{2}}\sum_{x^2=-\frac{L_2}{2}}^{\frac{L_2}{2}}\frac{g_0^2}{2a}[l_2(x)]^2-\sum_{x^1=-\frac{L_1}{2}}^{\frac{L_1}{2}}\sum_{x^2=-\frac{L_2}{2}}^{\frac{L_2}{2}}\frac{1}{2g_0^2 a}[{\rm Tr}\, U_2(x^1,x^2)^\dag U_2(x^1+a,x^2)+c.c.]\nonumber\\
&&-\frac{g_0^2}{2a}\sum_{x^1=-\frac{L_1}{2}}^{\frac{L_1}{2}}\sum_{y^1=-\frac{L_1}{2}}^{\frac{L_1}{2}}\sum_{x^2=-\frac{L_2}{2}}^{\frac{L_2}{2}}\vert x^1-y^1\vert[l_2(x^1,x^2)-\mathcal{R}_2(x^1,x^2-a)l_2(x^1,x^2-a)]\nonumber\\
&&\,\,\,\,\,\,\,\,\,\,\,\,\,\,\times[l_2(y^1,x^2)-\mathcal{R}_2(y^1,x^2-a)l_2(y^1,x^2-a)].\label{latticeafteraxial}
\eeq
The Hamiltonian (\ref{latticeafteraxial}) is nonlocal in $x^1$, and depends only on the transverse degrees of freedom $U_2,\,l_2$. 


We now explore anisotropic Yang-Mills theory by longitudinally rescaling the coordinates. This is a summary of the calculation done in \cite{twoplusone}, \cite{horizontal}. The longitudinally-rescaled lattice has spacing $\lambda a$ in the $x^{0,1}$ directions and spacing $a$ in the $x^2$-direction. In the $\lambda\to0$ limit, it is sensible to treat $x^0$ and $x^1$ as continuous directions, and $x^2$ discrete.

Longitudinally rescaling the lattice Hamiltonian ({\ref{latticeafteraxial}), gives $H=H_0+\lambda^2H_1$, where
\beq
H_0&=&\sum_{x^1=-\frac{L_1}{2}}^{\frac{L_1}{2}}\sum_{x^2=-\frac{L_2}{2}}^{\frac{L_2}{2}}\frac{g_0^2}{2a}[l_2(x)]^2-\sum_{x^1=-\frac{L_1}{2}}^{\frac{L_1}{2}}\sum_{x^2=-\frac{L_2}{2}}^{\frac{L_2}{2}}\frac{1}{2g_0^2 a}[{\rm Tr}\, U_2(x^1,x^2)^\dag U_2(x^1+a,x^2)+c.c.],\nonumber\\
H_1&=&-\frac{\lambda^2g_0^2}{2a}\sum_{x^1=-\frac{L_1}{2}}^{\frac{L_1}{2}}\sum_{y^1=-\frac{L_1}{2}}^{\frac{L_1}{2}}\sum_{x^2=-\frac{L_2}{2}}^{\frac{L_2}{2}}\vert x^1-y^1\vert\nonumber\\
&&\times[l_2(x^1,x^2)-\mathcal{R}_2(x^1,x^2-a)l_2(x^1,x^2-a)][l_2(y^1,x^2)-\mathcal{R}_2(y^1,x^2-a)l_2(y^1,x^2-a)].\nonumber
\eeq
Henceforth we drop the Lorentz index $2$ from $U_2$, $l_2$.

We treat $H_1$ as a perturbation. In the interaction representation, $U$ satisfies the Heisenberg equation of motion, $\partial_0 U=i[H_0,U]$. The solution of this equation of motion is
\beq
l(x^1,x^2)_b&=&\frac{ia}{g_0^2}{\rm Tr}\, t_b\partial_0U(x^1,x^2)U(x^1,x^2)^\dag,\nonumber\\
\mathcal{R}(x^1,x^2)_b^{\,\,c}l(x^1,x^2)_c&=&\frac{ia}{g_0^2}{\rm Tr}\,t_b U(x^1,x^2)^\dag\partial_0 U(x^1,x^2).\label{heisenbergevolution}
\eeq
Substituting (\ref{heisenbergevolution}) into $H_0$, and taking the continuum limit in the $x^1$  direction, we find
\beq
H_0&=&\sum_{x^2}H_0(x^2)=\sum_{x^2} \int dx^1\frac{1}{2g_0^2}\left\{\left[j_0^L(x^1,x^2)_b\right]^2+\left[j_1^L(x^1,x^2)_b\right]^2\right\}\nonumber\\
&=&\sum_{x^2} \int dx^1\frac{1}{2g_0^2}\left\{\left[j_0^R(x^1,x^2)_b\right]^2+\left[j_1^R(x^1,x^2)_b\right]^2\right\},\nonumber
\eeq
where
\beq
j_\mu^L(x)_b=i{\rm Tr}\, t_b \partial_\mu U(x) U(x)^\dag,\,\,j_\mu^R(x)_b=i{\rm Tr}\, t_b U(x)^\dag\partial_\mu U(x),\label{leftandright}
\eeq
where $\mu=0,1$.

We now note that $H_0(x^2)$ is the Hamiltonian of a (1+1)-dimensional PCSM located at $x^2$. The PCSM has the action
\beq
\mathcal{L}_{\rm PCSM}=\int d^2x \frac{1}{2g_0^2}\eta^{\mu\nu}{\rm Tr}\,\partial_\mu U^\dag \partial_\nu U.\label{actionpcsmone}
\eeq
This model has a global $SU(N)\times SU(N)$ symmetry given by the transformation $U(x)\to V^L U(x) V^R$, where $V^{L,R}\in SU(N)$. The Noether currents corresponding to these global symmetries are $j^{L,R}$ given in (\ref{leftandright}). The Hamiltonian corresponding to the action (\ref{actionpcsmone}) of a single PCSM at fixed $x^2$ is $H_0(x^2)$. The unperturbed Hamiltonian, $H_0$, is an array of PCSM's, one at each value of $x^2$,
\beq
H_0=\sum_{x^2}H_0(x^2)=\sum_{x^2}H_{\rm PCSM}(x^2).\nonumber
\eeq
It is important to note that the PCSM is known to be integrable and to have a mass gap. We call $m$ the mass of the elementary particles of the sigma model.

The residual Gauss's law, (\ref{residualgausslattice}) becomes
\beq
\int dx^1\left[ j_0^L(x^1,x^2)_b-j_0^R(x^1,x^2-a)_b\right]\Psi=0,\label{globalgausslaw}
\eeq
for each value of $x^2$, when $x^1$ is continuous.

Using (\ref{heisenbergevolution}), we write the interaction Hamiltonian $H_1$ in the continuous $x^1$ limit:
\beq
H_1&=&\sum_{x^2}\int dx^1\int dy^1\frac{1}{4g_0^2 a}\vert x^1-y^1\vert\left[j_0^L(x^1,x^2)-j_0^R(x^1,x^2-a)\right]\left[j_0^L(y^1,x^2)-j_0^R(y^1,x^2-a)\right].\label{perturbationhamiltonian}
\eeq
The Hamiltonian (\ref{perturbationhamiltonian}) couples adjacent sigma models, which allows particles to propagate in the $x^2$-direction. The coupling is suppressed in the $\lambda\to 0$ limit.

There are several important points to mention about the Hamiltonian $H=H_0+\lambda^2H_1$. It has been shown that this anisotropic model confines quarks. The string tensions are different if there is a quark-antiquark pair separated in the $x^1$ or the $x^2$-direction. We call these the horizontal string tension, $\sigma^H$ and the vertical string tension $\sigma^V$, respectively. To lowest order in $\lambda$, these are given by \cite{horizontal}, \cite{vertical},
\beq
\sigma^H=\lambda^2\frac{g_0^2}{a^2}C_N,\,\,\,\,\,\,\,\,\,\,\,\,\,\sigma^V=\frac{m}{a},\label{stringtensions}
\eeq
where $C_N$ is the smallest eigenvalue of the Casimir operator of $SU(N)$. In Sections III and IV, we compute quantum corrections to the string tensions (\ref{stringtensions}) using the exact form factors of the sigma model (shown in the appendix). This calculation is a generalization of the results computed by Orland in References \cite{horizontal} and \cite{vertical} for the gauge group $SU(2)$. Orland's results were computed using the form factors of the $O(4)$-symmetric nonlinear sigma model \cite{karowski}, by virtue of $SU(2)\times SU(2)\simeq O(4)$. Recently some form factors of the PCSM for general $N>2$ have been found \cite{multiparticle}, which allow us to generalize Orland's result to the gauge group $SU(N)$.

The anisotropic Hamiltonian has a mass gap. The lightest gauge invariant excitation is a glueball composed of a sigma-model particle-antiparticle pair. The light gluon mass spectrum was calculated by Orland for the gauge group SU(2) in Ref. \cite{glueball}. The glueball masses are of the form
\beq
M_n=2m+E_n,\nonumber
\eeq
where $E_n$ is the binding energy of the particle-antiparticle pair. The determination of the spectrum of energies, $E_n,$ involved knowledge of the exact S-matrix of the O(4) sigma model \cite{zamolodchikov}. We generalize this calculation for $N>2$ in Section V, using the exact S-matrix of the PCSM found by Wiegmann \cite{wiegmann}.

\section{The Horizontal String Tension}
\setcounter{equation}{0}
\renewcommand{\theequation}{3.\arabic{equation}}

In this section we compute quantum corrections to the string tension $\sigma^H$. This calculation has been done before, in Reference \cite{horizontal}, for $N=2$ using the form factors of the $O(4)$ sigma model. In this section we generalize these results for $N>2$.

It is convenient to rewrite the Hamiltonian (\ref{perturbationhamiltonian}) by reintroducing the auxiliary field $\Phi=-A_0$, such that
\beq
H_1=\sum_{x^2}\int dx^1\left\{\frac{g_0^2\, a^2}{4}\partial_1 \Phi(x^1,x^2)\partial_1 \Phi(x^1,x^2)-j_0^L(x^1,x^2)\Phi(x^1,x^2)-j_0^{R}(x^1,x^2)\Phi(x^1,x^2+a)\right\}.\label{withauxiliary}
\eeq
By integrating out the auxiliary field, $\Phi$, we see the Hamiltonians, (\ref{withauxiliary}) and (\ref{perturbationhamiltonian}) are equivalent.

We can easily introduce static quarks into the Hamiltonian (\ref{withauxiliary}) by coupling them to the auxiliary field, $\Phi$. Our goal is to find the potential energy of a quark-antiquark pair separated only in the $x^1$-direction. By integrating out the sigma model degrees of freedom, we can find the quantum corrections to the string tension $\sigma^H$. The Hamiltonian with a static quark of charge $q$ at the space point $(u^1,u^2)$, and an antiquark of charge $q^\prime$ at the space-time point $(v^1,v^2)$, is 
\beq
H_1&=&\sum_{x^2}\int dx^1\left\{\frac{g_0^2\, a^2}{4}\partial_1 \Phi(x^1,x^2)\partial_1 \Phi(x^1,x^2)-j_0^L(x^1,x^2)\Phi(x^1,x^2)-j_0^{R}(x^1,x^2)\Phi(x^1,x^2+a)\right.\nonumber\\
&&\,\,\,\,\,\,\,\,\,\,\,\,\,\,\,\,\,\,\,\,\,\left.+g_0^2\,q\,\Phi (u^1,u^2)-g_0^2\,q^\prime\Phi(v^1,v^2)\right\}.\label{withquarks}
\eeq
With these static quarks, the residual gauss law on physical states is modified to:
\beq
&&\int dx^1\left[j_0^L(x^1,x^2)_b-j_0^R(x^1,x^2-a)_b+q_b\delta(x^1-u^1)\delta_{x^2 u^2}-q^\prime_b\delta(x^1-v^1)\delta_{x^2 v^2}\right]\Psi=0.\label{modifiedresidualgauss}
\eeq

To find the string tension, $\sigma^H$, we set $u^2=v^2$, and integrate out the sigma model field, $U$. We obtain an effective action, $S_{\rm eff}(\Phi)$, by
\beq
e^{iS_{\rm eff}(\Phi)}=\langle 0\vert \mathcal{T}\, e^{i\int dx^0 \lambda^2 H_1}\vert 0\rangle,\label{integrateout}
\eeq
where $\mathcal{T}$ stands for time ordering. The field $\Phi$ in (\ref{integrateout}) is treated as a background classical field. Expanding (\ref{integrateout}) in powers of $\lambda$, up to quartic order, we find
\beq
S_{\rm eff}(A_0)\approx&& -i\lambda^2\sum_{x^2}\int d^2x\,\frac{g_0^2a^2}{4}\Phi\partial_1^2\Phi+i\lambda^4S^{(2)}(\Phi)+\mathcal{O}(\lambda^6)\nonumber\\
&&-\lambda^2\sum_{x^2}\int d^2x\left[ g_0^2\,q(x^0)\,\Phi (x^0,u^1,u^2)-g_0^2\,q^\prime(x^0)\Phi(x^0,v^1,v^2)\right]\!\!\!,\label{connectedgraph}
\eeq
where
\beq
iS^{(2)}\equiv-\frac{1}{2}\sum_{x^2}\int d^2x d^2y \,D(x^0, x^1,y^0,y^1,x^2)_{acef} \Phi(x^0,x^1,x^2)_{ac}\Phi(y^0,y^1,x^2)_{ef},\nonumber
\eeq
where
\beq 
D(x^0, x^1,y^0,y^1,x^2)_{acef}\equiv\langle 0\vert \mathcal{T} j_0^L(x^0,x^1,x^2)_{ac}\,j_0^L(y^0,y^1,x^2)_{ef}\vert 0\rangle.\label{twopointtimeordered}
\eeq

We compute the correlation function (\ref{twopointtimeordered}) by introducing a complete set of intermediate states between the two operators. The non-time-ordered correlation function is given by 
\beq
&&\langle 0\vert  j_0^L(x)_{ac}\,j_0^L(y)_{ef}\vert 0\rangle=\sum_{M=1}^{\infty}\frac{1}{N(M!)^2}\int\frac{d\theta_1\dots d\theta_{2M}}{(2\pi)^{2M}}e^{-i(x-y)\cdot\left[\sum_{j=1}^{2M} p_j\right]}\nonumber\\
&&\,\,\,\,\,\times\langle0|j_\mu^L(0)_{a_0c_0}|A,\theta_1,b_1,a_1;\dots;A,\theta_M,b_Ma_M;P,\theta_{M+1},a_{M+1},b_{M+1};\dots;P,\theta_{2M},a_{2M},b_{2M}\rangle\nonumber\\
&&\,\,\,\,\,\times\left[\langle0|j_\nu^L(0)_{e_0f_0}|A,\theta_1,b_1,a_1;\dots;A,\theta_M,b_Ma_M;P,\theta_{M+1},a_{M+1},b_{M+1};\dots;P,\theta_{2M},a_{2M},b_{2M}\rangle\right]^*.\nonumber
\eeq
The correlation function (\ref{twopointtimeordered}) can be found exactly at large $N$ using the form factors from Ref. \cite{myselforland}. For general $N<\infty$, we can only calculate a large-distance approximation, using the two-particle form factor (shown in the appendix). At large distances, it is sufficient to compute only the first intermediate state, with one particle and one antiparticle. 

The sigma-model form factor with one particle and one antiparticle is (see the appendix)
\beq
&&\langle 0\vert j_\mu^L(x)_{ac}\vert A,\theta_1,b_1,a_1;P,\theta_2,a_2,b_2\rangle=(p_1-p_2)_\mu\left( \delta_{a_0a_2}\delta_{c_0a_1}\delta_{b_1b_2}-\frac{1}{N}\delta_{a_0c_0}\delta_{a_1a_2}\delta_{b_1b_2}\right)e^{-ix\cdot (p_1+p_2)}\nonumber\\
&&\,\,\,\,\,\,\,\,\,\,\,\,\times\frac{2\pi i}{(\theta+\pi i)}\exp \int_0^\infty \frac{d\xi}{\xi}\left[\frac{-2\sinh\left(\frac{2\xi}{N}\right)}{\sinh \xi}+\frac{4e^{-\xi}\left(e^{2\xi/N}-1\right)}{1-e^{-2\xi}}\right]\frac{\sin^2[\xi(\pi i-\theta)/2\pi]}{\sinh \xi}.\label{twoparticlerecall}
\eeq
Inserting (\ref{twoparticlerecall}) into (\ref{twopointtimeordered}) and time ordering, we find
\beq
D(x,y)_{acef}&=&\int\frac{d\theta_1 \, d\theta_2}{(2\pi)^2}m^2(\cosh\theta_1-\cosh\theta_2)^2\left(\delta_{a a_2}\delta_{c a_1}-\frac{1}{N}\delta_{a c}\delta_{a_1 a_2}\right)\left(\delta_{e a_2}\delta_{f a_1}-\frac{1}{N}\delta_{e f}\delta_{a_1 a_2}\right)\nonumber\\
&&\times\exp\left\{-im\,{\rm sgn}(x^0-y^0)\left[(x^0-y^0)(\cosh\theta_1+\cosh\theta_2)-(x^1-y^1)(\sinh\theta_1+\sinh\theta_2)\right]\right\}\nonumber\\
&&\times\left\{\frac{2\pi i}{(\theta+\pi i)}\exp \int_0^\infty \frac{d\xi}{\xi}\left[\frac{-2\sinh\left(\frac{2\xi}{N}\right)}{\sinh \xi}+\frac{4e^{-\xi}\left(e^{2\xi/N}-1\right)}{1-e^{-2\xi}}\right]\frac{\sin^2[\xi(\pi i-\theta)/2\pi]}{\sinh \xi}\right\}^2.\label{timeorderedtwopoint}
\eeq
The color factor in (\ref{timeorderedtwopoint}) is
\beq
\left(\delta_{a a_2}\delta_{c a_1}-\frac{1}{N}\delta_{a c}\delta_{a_1 a_2}\right)\left(\delta_{e a_2}\delta_{f a_1}-\frac{1}{N}\delta_{e f}\delta_{a_1 a_2}\right)=\delta_{ae}\delta_{ef}-\frac{1}{N}\delta_{ac}\delta_{ef}.\label{deltas}
\eeq
The term in the right-hand side of (\ref{deltas}) proportional to $\frac{1}{N}$ does not contribute when we substitute (\ref{timeorderedtwopoint}) back into (\ref{connectedgraph}), because the field $\Phi$ is traceless, so we will ignore this term from now on.

We evaluate $iS^{(2)}(\Phi)$ using coordinates $X^\mu,\,r^\mu$, defined by $x^\mu=X^\mu+\frac{1}{2}r^\mu,$ and $y^\mu=X^\mu-\frac{1}{2}r^\mu$. We then use the derivative expansion for $X\gg r$:
\beq
\Phi(x)&=&\Phi(X)+\frac{r^\mu}{2}\partial_\mu \Phi(X)+\frac{r^\mu r^\nu}{8}\partial_\mu\partial_\nu \Phi(X)+\dots,\nonumber\\
\Phi(y)&=&\Phi(X)-\frac{r^\mu}{2}\partial_\mu \Phi(X)+\frac{r^\mu r^\nu}{8}\partial_\mu \partial_\nu \Phi(X)\pm\dots,\label{derivativeexpansion}
\eeq
where $\partial_\mu$ denotes $\partial/\partial X^\mu$. This derivative expansion is valid at large distances. The quadratic contribution to the effective action is 
\beq
iS^{(2)}&=&-\frac{i}{2}\int d^2Xd^2r\,D\left(X+\frac{r}{2},X-\frac{r}{2}\right)_{acef}\Phi\left(X+\frac{r}{2}\right)_{ac}\,\Phi\left(X-\frac{r}{2}\right)_{ef}.\label{integrater}
\eeq

We substitute (\ref{derivativeexpansion}) into (\ref{integrater}) and find
\beq
iS^{(2)}&=&-\frac{i}{2}\int d^2Xd^2r\,\int\frac{d\theta_1 \, d\theta_2}{(2\pi)^2}m^2(\cosh\theta_1-\cosh\theta_2)^2 \delta_{a e}\delta_{c f}\nonumber\\
&&\times\exp\left\{-im\,{\rm sgn}(r^0)\left[(r^0)(\cosh\theta_1+\cosh\theta_2)-(r^1)(\sinh\theta_1+\sinh\theta_2)\right]\right\}\nonumber\\
&&\times\left\{\frac{2\pi i}{(\theta+\pi i)}\exp \int_0^\infty \frac{d\xi}{\xi}\left[\frac{-2\sinh\left(\frac{2\xi}{N}\right)}{\sinh \xi}+\frac{4e^{-\xi}\left(e^{2\xi/N}-1\right)}{1-e^{-2\xi}}\right]\frac{\sin^2[\xi(\pi i-\theta)/2\pi]}{\sinh \xi}\right\}^2\nonumber\\
&&\times \left(\Phi(X)_{ac}+\frac{r^\mu}{2}\partial_\mu \Phi(X)_{ac}+\frac{r^\mu r^\nu}{8}\partial_\mu\partial_\nu \Phi(X)_{ac}\right)\left(\Phi(X)_{ef}-\frac{r^\mu}{2}\partial_\mu \Phi(X)_{ef}+\frac{r^\mu r^\nu}{8}\partial_\mu \partial_\nu \Phi(X)_{ef}\right)\!\!.\label{explicitintegrate}
\eeq
We keep only terms quadratic in $r$ in (\ref{explicitintegrate}) and then integrate out the $r$ variable. Only the terms proportional to $(r^1)^2$ give a non-vanishing contribution in (\ref{explicitintegrate}). 
Integration yields the effective action:
\beq
S_{\rm eff}(\Phi)&=&\int d^2x\frac{1}{2}\Phi\partial_1^2\Phi\left\{1-\lambda^2\frac{Nm}{2(2\pi)^2}\int d\theta_1 d\theta_2 \frac{\sinh^2\left(\frac{\theta_1+\theta_2}{2}\right)\sinh^2\left(\frac{\theta_1-\theta_2}{2}\right)}{\cosh \left(\frac{\theta_1+\theta_2}{2}\right)\cosh\left(\frac{\theta_1-\theta_2}{2}\right)}\right.\nonumber\\
&&\times\left.\delta^{\prime\prime}\left[2m\cosh\left(\frac{\theta_1+\theta_2}{2}\right)\sinh\left(\frac{\theta_1-\theta_2}{2}\right)\right]\frac{4\pi^2}{\left(\theta_1-\theta_2\right)^2+\pi^2}\right.\nonumber\\
&&\left.\times\exp 2\int_0^\infty \frac{d\xi}{\xi}\left[\frac{-2\sinh\left(\frac{2\xi}{N}\right)}{\sinh \xi}+\frac{4e^{-\xi}\left(e^{2\xi/N}-1\right)}{1-e^{-2\xi}}\right]\frac{\sin^2[\xi(\pi i-(\theta_1-\theta_2))/2\pi]}{\sinh \xi}\right\}\nonumber\\
&-&\lambda^2\sum_{x^2}\int d^2x\left[ g_0^2\,q(x^0)\,\Phi (x^0,u^1,u^2)-g_0^2\,q^\prime(x^0)\Phi(x^0,v^1,v^2)\right].\nonumber
\eeq

We read off the renormalized string tension $\sigma^H$, after integrating out the auxiliary field $\Phi$:
\beq
\sigma^H&=&\lambda^2\, \frac{g_0^2}{a^2}C_N\left\{1-\left[\lambda^2\frac{Nm}{2(2\pi)^2}\int d\theta_1 d\theta_2 \frac{\sinh^2\left(\frac{\theta_1+\theta_2}{2}\right)\sinh^2\left(\frac{\theta_1-\theta_2}{2}\right)}{\cosh \left(\frac{\theta_1+\theta_2}{2}\right)\cosh\left(\frac{\theta_1-\theta_2}{2}\right)}\right.\right.\nonumber\\
&&\,\,\,\,\,\,\,\,\,\,\,\,\,\,\,\,\,\,\,\,\left.\times\delta^{\prime\prime}\left(2m\cosh\left(\frac{\theta_1+\theta_2}{2}\right)\sinh\left(\frac{\theta_1-\theta_2}{2}\right)\right)\frac{4\pi^2}{\left(\theta_1-\theta_2\right)^2+\pi^2}\right.\nonumber\\
&&\,\,\,\,\,\,\,\,\,\,\,\,\,\,\,\,\,\,\,\,\,\left.\times\exp 2\int_0^\infty \frac{d\xi}{\xi}\left[\frac{-2\sinh\left(\frac{2\xi}{N}\right)}{\sinh \xi}\right.\left.\left.+\frac{4e^{-\xi}\left(e^{2\xi/N}-1\right)}{1-e^{-2\xi}}\right]\frac{\sin^2[\xi(\pi i-(\theta_1-\theta_2))/2\pi]}{\sinh \xi}\right]\right\}^{-1}\nonumber.
\eeq
After the integration over $\theta_1$ and $\theta_2$, the string tension is
\beq
\sigma^H=\lambda^2\, \frac{g_0^2}{a^2}C_N\left\{1-\lambda^2\frac{N}{3m^3 (2\pi)^2}\exp2\int_0^\infty\frac{d\xi}{\xi}\left[\frac{-2\sinh\left(\frac{2\xi}{N}\right)}{\sinh \xi}+\frac{4e^{-\xi}\left(e^{2\xi/N}-1\right)}{1-e^{-2\xi}}\right]\frac{\sin^2\left(\frac{i\xi}{2}\right)}{\sinh\xi}\right\}^{-1}.\label{horizontaltension}
\eeq
The string tension (\ref{horizontaltension}) generalizes the result from Refence \cite{horizontal} from $N=2$, to general $N>2$. 

In the next section we compute the string tension for a quark-antiquark pair separated in the $x^2$-direction, rather than the $x^1$-direction. 

\section{The Vertical String Tension}
\setcounter{equation}{0}
\renewcommand{\theequation}{4.\arabic{equation}}

In this section we calculate the string tension, $\sigma^V$, for a quark-antiquark pair separated only in the $x^2$-direction. This calculation has been done before in Reference \cite{vertical} for the $SU(2)$ gauge group. We show here how to generalize this result for $N>2$ using the form factors from Reference \cite{multiparticle}.

If we place a static quark at the space point $u^1,u^2$, and an antiquark at $u^1,v^2$, with $u^2>v^2$, The residual Gauss's Law (\ref{modifiedresidualgauss}) requires that there be at least one sigma-model particle in each $x^2$ layer, for $u^2>x^2>v^2$. The left-handed color index of a particle at $x^2$ is contracted with the right-handed color of the particle at $x^2+a$. The left-handed color index of the particle at $u^2-a$ and the right-handed color of the particle at $v^2+a$ are contracted with the color indices of the quark at $u^2$, and the antiquark at $v^2$, respectively. The physical state  is a color-singlet string of sigma-model particles, whose endpoints are the quarks.
The vertical string tension is obtained by calculating the energy of this string,
\beq
\sigma^V=\lim_{\vert u^2-v^2\vert\to\infty} \frac{E_{\rm string}}{\vert u^2-v^2\vert}.\nonumber
\eeq
The first approximation is to assume the energy of the string is the total mass of the sigma-model particles, such that $E_{\rm string}=\frac{m}{a}\vert u^2-v^2\vert$, so $\sigma^V=m/a$.

Corrections to the vertical string tension are found by calculating the contributions to the energy of the string from the Hamiltonian $\lambda^2H_1$. As in Reference \cite{vertical}, we will use a nonrelativistic approximation, where the sigma-model particles have momenta much smaller than their masses. We ignore any creation or annihilation of particles.

The projection of the Hamiltonian onto the nonrelativistic string state is
\beq
H=\sum_{x^2=v^2}^{u^2}\left\{m+\int\frac{dp}{2\pi}\frac{p^2}{2m}\mathfrak{A}_P^\dag(p)_{a b}\mathfrak{A}_P(p)_{a b}\right\}+\lambda^2H_1,\nonumber
\eeq
where  $\mathfrak{A}_P^\dag(p)_{a b}$, and $\mathfrak{A}_P(p)_{a b}$ are the sigma-model particle creation and annihilation operators, respectively, and 
\beq
&&H_1=\sum_{x^2}\int dx^1\int dy^1\frac{1}{4g_0^2 a}\vert x^1-y^1\vert\nonumber\\
&&\,\,\,\,\times\left[j_0^L(x^1,x^2)-j_0^R(x^1,x^2-a)+q_b\delta(x^1-u^1)\delta_{x^2 u^2}-q^\prime_b\delta(x^1-u^1)\delta_{x^2 v^2}\right]\nonumber\\
&&\,\,\,\,\times\left[j_0^L(y^1,x^2)-j_0^R(y^1,x^2-a)+q_b\delta(y^1-u^1)\delta_{x^2 u^2}-q^\prime_b\delta(y^1-u^1)\delta_{x^2 v^2}\right],\nonumber\\
\label{stringhamiltonian}
\eeq
where we have again eliminated the auxiliary field, $A_0$.

We now need to find the expectation value 
\beq
\langle {\rm string}\vert H_1\vert {\rm string}\rangle,\label{expectationstring}
\eeq
 where the state $\vert {\rm string}\rangle$ has a sigma-model particle for every $x^2$, whose center of mass is located at $x^1=z(x^2)$. To evaluate (\ref{expectationstring}), we need the matrix elements of the form
 \beq
 &&\langle P, z_1,a_1,b_1\vert j_0^{C}(x)_{ac}\vert P,z_2,a_2,b_2\rangle=\int \frac{dp_1}{2\pi}\frac{1}{\sqrt{2E_1}}\int\frac{dp_2}{2\pi}\frac{1}{\sqrt{2E_2}}\nonumber\\
 &&\,\,\,\,\,\,\,\,\times e^{-ip_1\cdot(z_1-x)+ip_2\cdot(z_2-x)}\langle P, \theta_1,a_1,b_1\vert j_0^C(x)_{ac}\vert P,\theta_2,a_2,b_2\rangle,\label{fourierform}
 \eeq
where the matrix element on the right hand side of (\ref{fourierform}) is the two particle form factor found in the appendix (with the incoming antiparticle crossed to an outgoing particle), and $C=L,R$. By applying crossing symmetry on the form factor (\ref{finitenform}), we find
\beq
&&\langle P, \theta_1,a_1,b_1\vert j_0^C(x)_{ac}\vert P,\theta_2,a_2,b_2\rangle\nonumber\\
&&\,\,\,\,=(p_1+p_2)_0\mathfrak{D}^C_{a\,c\,a_1a_2b_1b_2}\frac{2\pi i}{\theta+2\pi i}\exp\int_0^\infty\frac{d\xi}{\xi}\left[\frac{-2\sinh\left(\frac{2\xi}{N}\right)}{\sinh\xi}+\frac{4 e^{-\xi}\left(e^{\frac{2\xi}{N}}-1\right)}{1-e^{-2\xi}}\right]\frac{\sin^2[\xi\theta/2\pi]}{\sinh\xi},\nonumber
\eeq
where
\beq
\mathfrak{D}^L_{a\,c\,a_1a_2b_1b_2}&=&\delta_{a\,a_2}\delta_{c\,a_1}\delta_{b_1b_2}-\frac{1}{N}\delta_{a\,c\,}\delta_{a_1a2}\delta_{b_1b_2},\nonumber\\
\mathfrak{D}^R_{a\,c\,a_1a_2b_1b_2}&=&\delta_{a\,b_2}\delta_{c\,b_1}\delta_{a_1a_2}-\frac{1}{N}\delta_{a\,c}\delta_{a_1a_2}\delta_{b_1b_2}.\nonumber
\eeq
Taking the nonrelativistic limit, we find
\beq
\frac{1}{\sqrt{2E_1}}\frac{1}{\sqrt{2E_2}}\langle P, \theta_1,a_1,b_1\vert j_0^C(x)_{a\,c}\vert P,\theta_2,a_2,b_2\rangle\approx\mathfrak{D}^C_{a\,c\,a_1a_2b_1b_2}\exp-\frac{A_N}{m^2}(p_1-p_2)^2.\nonumber
\eeq
where
\beq
A_N=\int_0^\infty\frac{d\xi}{4\pi^2}\frac{\xi}{\sinh\xi}\left[\sinh\left(\frac{2\xi}{N}\right)-2\left(e^{2\xi/N}-1\right)\right]=\frac{1}{16}\pi^2\left[2\pi^2-3\psi^{(1)}\left(\frac{1}{2}-\frac{1}{N}\right)-\psi^{(1)}\left(\frac{1}{2}+\frac{1}{N}\right)\right],\nonumber
\eeq
for $N>2$, where $\psi^{(n)}(x)=d^{n+1}\ln \Gamma(x)/dx^{n+1}$ is the $n$-th polygamma function.

The matrix element $(\ref{fourierform})$ is then
\beq
\langle P, z_1,a_1,b_1\vert j_0^{C}(x)_{ac}\vert P,z_2,a_2,b_2\rangle=\sqrt{\frac{m^2}{2\pi A_N}}\mathfrak{D}^C_{a\,c\,a_1a_2b_1b_2}\exp\left[-\frac{m^2}{4A_N}\left(\frac{z_1+z_2}{2}-y\right)^2\right]\delta(z_1-z_2).\nonumber\\\label{smearedcolor}
\eeq
This means that the color density of a particle is a Gaussian distribution in the nonrelativistic limit. In this sense, particles are not point-like, but the color is smeared over space.

We now use (\ref{smearedcolor}) to write the effective Hamiltonian of the nonrelativistic string. This is given by the projection of the Hamiltonian (\ref{stringhamiltonian}) onto the state $\vert {\rm string}\rangle$, which has a sigma-model particle at each $x^2$ layer, located at the point $z^{1}(x^2)$, for  $u^2>x^2>v^2$, a static quark at $u^1,u^2$, and an antiquark at $u^1,v^2$. The string Hamiltonian is
\beq
H_{\rm string}=\frac{m}{a}(v^2-u^2)-\frac{1}{2m}\sum_{x^2=v^2}^{u^2-a}\frac{\partial^2}{\partial z^1(x^2)^2}+\lambda^2V_{\rm bulk}+\lambda^2V_{\rm ends},\nonumber
\eeq
where
\beq
V_{\rm bulk}&=&-\frac{m^2}{8\pi A_N}\frac{1}{g_0^2 a^2}\sum_{x^2=v^2+a}^{u^2-a}\int dx^1\,dy^1\vert x^1-y^1\vert\nonumber\\
&&\times\left\{e^{-\frac{m^2}{4A_N}\left[z^1(x^2)-x^1\right]^2}\mathfrak{D}^L(x^2)_{a\,c\,a_1a_2b_1b_2}-e^{-\frac{m^2}{4A_N}\left[z^1(x^2-a)-x^1\right]^2}\mathfrak{D}^R(x^2-a)_{a\,c\,a_1a_2b_1b_2}\right\}\nonumber\\
&&\times\left\{e^{-\frac{m^2}{4A_N}\left[z^1(x^2)-y^1\right]^2}\mathfrak{D}^L(x^2)_{a\,c\,a_2a_1b_2b_1}-e^{-\frac{m^2}{4A_N}\left[z^1(x^2-a)-y^1\right]^2}\mathfrak{D}^R(x^2-a)_{a\,c\,a_2a_1b_2b_1}\right\}\label{bulk},
\eeq
and
\beq
V_{\rm ends}&=&-\frac{1}{4g_0^2 a^2}\int dx^1 dy^1\vert x^1-y^1\vert\left\{\sqrt{\frac{m^2}{2\pi A_N}}e^{-\frac{m^2}{4A_N}\left[z^1(v^2)-x^1\right]^2}\mathfrak{D}^R(v^2)_{a\,c\,a_1a_2b_1b_2}+\delta(x^2-v^1)q^\prime_{a\,c}4\pi\delta_{a_1a_2}\delta_{b_1b_2}\right\}\nonumber\\
&&\times\left\{\sqrt{\frac{m^2}{2\pi A_N}}e^{-\frac{m^2}{4A_N}\left[z^1(v^2)-y^1\right]^2}\mathfrak{D}^R(v^2)_{a\,c\,a_1a_2b_1b_2}+\delta(y^1-u^1)q^\prime_{a\,c}4\pi\delta_{a_1a_2}\delta_{b_1b_2}\right\}\nonumber\\
&&-\frac{1}{4g_0^2 a^2}\int dx^1 dy^1\vert x^1-y^1\vert\left\{\sqrt{\frac{m^2}{2\pi A_N}}e^{-\frac{m^2}{4A_N}\left[z^1(u^2-a)-x^1\right]^2}\mathfrak{D}^L(u^2-a)_{a\,c\,a_1a_2b_1b_2}+\delta(x^2-u^1)q^\prime_{a\,c}4\pi\delta_{a_1a_2}\delta_{b_1b_2}\right\}\nonumber\\
&&\times\left\{\sqrt{\frac{m^2}{2\pi A_N}}e^{-\frac{m^2}{4A_N}\left[z^1(u^2-a)-y^1\right]^2}\mathfrak{D}^L(u^2-a)_{a\,c\,a_1a_2b_1b_2}+\delta(y^2-u^1)q^\prime_{a\,c}4\pi\delta_{a_1a_2}\delta_{b_1b_2}\right\}\label{ends}.
\eeq

Imposing the residual Gauss's law (\ref{modifiedresidualgauss}) on (\ref{bulk}) and (\ref{ends}), implies
\beq
&&\int dx^1\left\{-\sqrt{\frac{m^2}{2\pi A_N}}e^{-\frac{m^2}{4A_N}\left[z^1(x^2)-x^1\right]^2}\mathfrak{D}^L(x^2)_{a\,c\,a_1a_2b_1b_2}\right.\nonumber\\
&&\left.+\sqrt{\frac{m^2}{2\pi A_N}}e^{-\frac{m^2}{4A_N}\left[z^1(x^2-a)-x^1\right]^2}\mathfrak{D}^R(x^2-a)_{a\,c\,a_1a_2b_1b_2}\right\}\Psi=0,\label{gaussconstraintone}
\eeq
for $u^2>x^2>v^2$, and
\beq
&&\int dx^1\sqrt{\frac{m^2}{2\pi A_N}}\left\{e^{-\frac{m^2}{4A_N}\left[z^1(v^2)-x^1\right]^2}\mathfrak{D}^R(v^2)_{a\,c\,a_1a_2b_1b_2}-q^\prime_{a\,c}\delta(x^1-u^1)4\pi 
\delta_{a_1a_2}\delta_{b_1b_2}\right\}\Psi=0,\nonumber\\
&&\int dx^1\sqrt{\frac{m^2}{2\pi A_N}}\left\{e^{-\frac{m^2}{4A_N}\left[z^1(u^2-a)-x^1\right]^2}\mathfrak{D}^L(u^2-a)_{a\,c\,a_1a_2b_1b_2}-q_{a\,c}\delta(x^1-u^1)4\pi 
\delta_{a_1a_2}\delta_{b_1b_2}\right\}\Psi=0,\label{gaussconstrainttwo}
\eeq
respectively. The constraint (\ref{gaussconstraintone}) is satisfied by identifying $\mathfrak{D}^L(x^2)_{a\,c\,a_1a_2b_1b_2}=\mathfrak{D}^R(x^2-a)_{a\,c\,a_1a_2b_1b_2}$. The constraint (\ref{gaussconstrainttwo}) is satisfied by identifying $\mathfrak{D}^R(v^2)_{a\,c\,a_1a_2b_1b_2}=q^\prime_{a\,c}4\pi\delta_{a_1a_2}\delta_{b_1b_2}$, and $\mathfrak{D}^L(u^2-a)_{a\,c\,a_1a_2b_1b_2}=q_{a\,c}4\pi\delta_{a_1a_2}\delta_{b_1b_2}$. Using this, we can eliminate the color degrees of freedom from (\ref{bulk}) and (\ref{ends}).

Next we integrate out the variables $x^1$ and $y^1$ from equations (\ref{bulk}) and (\ref{ends}). The integrals involved are:
\beq
\int dx^1 dy^1\vert x^1-y^1\vert e^{-\frac{m^2}{4A_N}\left[(x^1)^2+(y^1)^2\right]}&=&\frac{4\sqrt{2\pi}A_N^{3/2}}{m^3},\nonumber\\
\int dx^1 dy^1\vert x^1-y^1\vert e^{-\frac{m^2}{4A_N}\left[(x^1+r)^2+(y^1)^2\right]}&=&\frac{4\sqrt{2\pi}A_N^{3/2}}{m^3}P(r),\nonumber\\
\int dx^1 \vert x^1-u^1\vert e^{-\frac{m^2}{4A_N}\left[x^1-z^1(U^2)\right]^2}&=&\frac{2A_N}{m^2}P\left[\sqrt{2}z^1(u^2)-\sqrt{2}u^1\right],\nonumber
\eeq
Where $P(r)$ is a function for which we do not have an exact analytic expression, but its behavior for small and large $r$ is
\beq
P(r)=\left\{\begin{array}{cc}
1+\frac{m^2r^2}{4A_N},&r<<\frac{1}{m},\\
\sqrt{\frac{\pi}{2A_N}}m\vert r\vert,& r>>\frac{1}{m}.\end{array}\right.\label{limitspfunction}
\eeq

After integrating out $x^1$, and $y^1$, the string Hamiltonian is
\beq
H_{\rm string}&=&\frac{m}{a}(u^2-v^2)-\frac{1}{2m}\sum_{x^2=v^2}^{u^2-a}\frac{\partial^2}{\partial z^1(x^2)^2}\nonumber\\
&&-\frac{\lambda^2N(N^2-1)}{m\,g_0^2 a^2}\sqrt{\frac{A_N}{2\pi}}\sum_{x^2=v^2+a}^{u^2}\left\{1-P\left[z^1(x^2)-z^1(x^2-a)\right]\right\}\nonumber\\
&&-\frac{\lambda^2N(N^2-1)}{m\,g_0^2 a^2}\sqrt{\frac{A_N}{2\pi}}\left(1+P\left\{\sqrt{2}\left[z^1(v^2)-u^1\right]\right\}+P\left\{\sqrt{2}\left[z^1(u^2-a)-u^1\right]\right\}\right),\label{afterintegratex}
\eeq
where we have used
\beq
\left(\mathfrak{D}^{C}\right)^2=N\left(N^2-1\right).\nonumber
\eeq
The potential energy between a static quark-antiquark pair is then determined by finding the ground state of the Hamiltonian (\ref{afterintegratex}).

We further simplify the Hamiltonian (\ref{afterintegratex}) using the small-gradient approximation. That is, in the nonrelativistic limit (when the sigma model mass gap is taken to be very large), we expect that the sigma-model particles in two adjacent $x^2$ layers are close to each other in the $x^1$-direction. Specifically, we assume $\vert z^1(x^2)-z^1(x^2-a)\vert<< m^{-1}$. At the endpoints of the string, we also assume $\vert z^1(v^2)-u^1\vert<< m^{-1}$, and $\vert z^1(u^2-a)-u^1\vert<< m^{-1}$.
Using Eq. (\ref{limitspfunction}), the small-gradient approximation gives the Hamiltonian
\beq
H_{\rm string}&=&\frac{\lambda^2N(N^2-1)}{m\,g_0^2 a^2}\sqrt{\frac{A_N}{2\pi}}+\frac{m}{a}(u^2-v^2)-\frac{1}{2m}\sum_{x^2=v^2}^{v^2-a}\frac{\partial^2}{\partial z^1(x^2)^2}\nonumber\\
&&+\frac{\lambda^2N(N^2-1)}{4 m g_0^2 a^2}\sqrt{\frac{1}{2\pi A_N}}\sum_{x^2=v^2+a}^{u^2-a}\left[z^1(x^2)-z^1(x^2-a)\right]^2\nonumber\\
&&+\frac{\lambda^2N(N^2-1)}{2 m g_0^2 a^2}\sqrt{\frac{1}{2\pi A_N}}\left\{\left[z^1(v^2)-u^1\right]^2+\left[z^1(u^2-a)-u^1\right]^2\right\}.\nonumber\\\label{smallgradientapprox}
\eeq

The first term in the Hamiltonian (\ref{smallgradientapprox}) is just a constant with no physical significance, so we will ignore it from now on. The Hamiltonian (\ref{smallgradientapprox}) is equivalent $Q=\left(u^2-v^2\right)/a$ coupled harmonic oscillators. The ground-state energy is then given by
\beq
E_0=mQ-\frac{\lambda\sqrt{N(N^2-1)}}{g_0a}\left(\frac{1}{2\pi A_N}\right)^{\frac{1}{4}}\sum_{q=0}^Q\sin\frac{\pi q}{2Q}.\label{groundstatestring}
\eeq
Using the Euler summation formula, for large $Q$:
\beq
\sum_{q=0}^QF\left(\frac{q}{Q}\right)&=&Q\int_0^1 dxF(x)-\frac{1}{2}\left[F(1)-F(0)\right]+\frac{1}{12Q}\left[F^\prime(1)-F^\prime(0)\right]+\mathcal{O}\left(\frac{1}{Q^2}\right), \nonumber
\eeq
the ground-state energy (\ref{groundstatestring}) becomes (dropping any constants that do not depend on $Q$)
\beq
E_0&=&\left[\frac{m}{a}-\frac{2\lambda\sqrt{N(N^2-1)}}{\pi g_0a^2}\left(\frac{1}{2\pi A_N}\right)^{\frac{1}{4}}\right]L+\frac{\pi}{24}\frac{\lambda\sqrt{N(N^2-1)}}{ g_0}\left(\frac{1}{2\pi A_N}\right)^{\frac{1}{4}}\frac{1}{L}+\mathcal{O}\left(\frac{1}{L^2}\right).\label{transversegroundstate}
\eeq
where the distance between the quark and antiquark is $L=Qa$. 

We can easily read the vertical string tension off (\ref{transversegroundstate}):
\beq
\sigma^V=\frac{m}{a}-\frac{2\lambda\sqrt{N(N^2-1)}}{\pi g_0a^2}\left(\frac{1}{2\pi A_N}\right)^{\frac{1}{4}}.\label{verticalstringtension}\eeq
There is also a Coulomb-like term in the quark-antiquark potential, which is proportional to $1/L$.

\section{The Low-lying Glueball Mass Spectrum}
\setcounter{equation}{0}
\renewcommand{\theequation}{5.\arabic{equation}}

The constraint (\ref{globalgausslaw}) requires that in the absence of quarks, there be an equal number of sigma-model particles and antiparticles in each $x^2$ layer. Furthermore, it requires that the excitations form left- and right-color singlets. If the sigma model at $x^2$ has a particle with a left color index, $a_1$, this index has to be  contracted with either the left-color index of an antiparticle in the $x^2$ layer, or the right color index of a particle in the $(x^2+a)$ layer. A glueball in this theory consists of several sigma-model excitations, forming a color-singlet bound state.

The simplest and lightest glueball is one composed of only one particle and one antiparticle, at the same value of $x^2$. The gauss law constraint requires that their left and right handed color indices be contracted. The interaction Hamiltonian (\ref{perturbationhamiltonian}) provides a confining linear potential, with string tension
\beq
\sigma=2\sigma^H.\label{stringhorizontal}
\eeq
The factor of $2$ comes the fact that both the left and right color charges are confined.

The problem is now essentially (1+1)-dimensional. The low-lying gluon spectrum has been found before by P. Orland in Reference \cite{glueball} for the $SU(2)$ gauge group. A similar analysis was used to find the massive spectrum of (1+1)-dimensional massive Yang-Mills theory for all $N$, in Ref. \cite{oneplusone}. This method is in turn inspired by the determination of the spectrum of the two-dimensional Ising model in an external magnetic field \cite{mccoy}, \cite{bhaseen}.

The low-lying glueball masses are 
\beq
M_n=2m+E_{n},\nonumber
\eeq
where $m$ is the mass of a sigma model excitation, and $E_n$ is the binding energy. The goal of this section is to compute the binding energies $E_n$ for $N>2$. This is done by finding the wave function of an unbound sigma-model particle-antiparticle pair. There is a possibility of these two excitations scattering which is accounted by the exact particle-antiparticle S-matrix. We later find the wave function of a particle-antiparticle pair, confined by a linear potential. We obtain a quantization condition for the binding energy by requiring that the two wave functions agree when the particles are close to each other. We are able to do this calculation only in the nonrelativistic limit, where we take the momenta of the excitations to be much smaller than their masses.

The particle-antiparticle S-matrix  is found in the appendix. The S-matrix has an incoming antiparticle with rapidity $\theta_1$ and color indices $a_1,b_1$ and a particle with rapidity $\theta_2$ and color indices $a_2,b_2$. There is an outgoing antiparticle with color indices $c_1,d_1$ and a particle with indices $c_2d_2$. The S-matrix is
\beq
S(\theta)_{a_1b_1;b_2a_2}^{d_2c_2;c_1d_1}=S(\theta)\left[\delta_{a_1}^{c_1}\delta_{a_2}^{c_2}-\frac{2\pi {\rm i}}{N(\pi {\rm i}-\theta)}\delta_{a_1a_2}\delta^{c_1c_2}\right]\left[\delta_{b_1}^{d_1}\delta_{b_2}^{d_2}-\frac{2\pi {\rm i}}{N(\pi {\rm i}-\theta)}\delta_{b_1b_2}b^{d_1d_2}\right],\nonumber
\eeq
where
\beq
S(\theta)=\exp2 \int_0^\infty\,\frac{d\xi}{\xi \sinh\xi}\left[ 2(e^{2\xi/N}-1)-\sinh(2\xi/N)
\right]\sinh\frac{\xi \theta}{\pi{\rm i}} \;,
\label{Stheta}
\eeq
for $N>2$.

The constraint (\ref{globalgausslaw}) requires that the particle-antiparticle pair form a left- and right-handed color singlet. The S-matrix of this pair, $\mathcal{S}(\theta)$, is obtained by contracting the color indices of the excitations:
\beq
\mathcal{S}(\theta)=\frac{1}{N^2}\delta_{a_1a_2}\delta_{b_1b_2}\delta_{c_1c_2}\delta_{d_1d_2}S(\theta)_{a_1b_1;b_2a_2}^{d_2c_2;c_1d_1}=\left(\frac{\theta+\pi i}{\theta-\pi i}\right)^2S(\theta).\nonumber
\eeq

In the nonrelativistic limit ($\theta<<m$) the color-singlet S-matrix becomes
\beq
\mathcal{S}(\theta)=\exp\left(-\frac{ih_N}{\pi m}\vert p_1-p_2\vert\right),\nonumber
\eeq
where
\beq
h_{N}&=&2\int_0^\infty \frac{d\xi}{\sinh \xi}\left[2(e^{2\xi/N}-1)-\sinh(2\xi/N)\right]\nonumber\\
&=&-4\gamma-\psi\left(\frac{1}{2}+\frac{1}{N}\right)-3\psi\left(\frac{1}{2}-\frac{1}{N}\right)-4\ln 4,
\label{N>2}
\eeq
and $\gamma$ is the Euler-Mascheroni constant, and 
$\psi(x)={d}\ln \Gamma(x)/dx$ is the digamma function.

We  find the wave function of an antiparticle at $x^1$, and a particle at $y^1$, with momenta $p_1,\,p_2$, respectively, in the nonrelativistic limit. It is convenient to switch to center-of-mass coordinates, $X,x$ and their respective momenta $P,p$. These are defined by $X=x^1+y^1$, $y^1-x^1$, $P=p_1+p_2$, and $p=p_2-p_1$. The nonrelativistic wave function is
\beq
\Psi_p(x)_{\rm singlet}=\left\{\begin{array}{c}
\cos(px+\omega),\,\,\,\,\,\,\,\,\,\,\,{\rm for}\,\,x>0,\\
\,\\
\cos[-px+\omega-\chi(p)],\,\,\,\,\,\,\,{\rm for}\,\,x<0,\end{array}\right.\label{singletstate}
\eeq
where $\chi(p)=-\frac{h_N}{\pi m}\vert p\vert.$

We now calculate the nonrelativistic wave function for a linearly-bound particle-antiparticle pair. In center-of-mass coordinates, the wave function satisfies the Schroedinger equation
\beq
-\frac{1}{m}\frac{d^2}{dx^2}\Psi(x)+\sigma \left\vert x \right\vert \,\Psi(x)=E\Psi(x),\label{schroedinger}
\eeq
where $E$ is the binding energy \cite{mccoy}, and $\sigma=2\lambda^2\, \frac{g_0^2}{a^2}C_N,$ is the string tension. The solution of Eq. (\ref{schroedinger}) is 
\beq
\Psi(x)=C{\rm Ai}\left[(m\sigma)^{\frac{1}{3}}\left(\vert x\vert -\frac{E}{\sigma}\right)\right],\label{airy}
\eeq
where ${\rm Ai}(x)$ is the Airy function of the first kind, and $C$ is a normalization constant.

We  require that the wave functions (\ref{singletstate}) and (\ref{airy}) agree as $\vert x\vert\to0$. We identify $\vert p\vert=(mE)^\frac{1}{2}$. For small $\vert x\vert$, the function (\ref{airy}) is approximated by
\beq
\Psi(x)_{b_1b_2}=\left\{\begin{array}{c}
C\frac{1}{\left(x+\frac{E}{\sigma}\right)^{\frac{1}{4}}}\cos \left[\frac{2}{3}(m\sigma)^{\frac{1}{2}}\left(-x+\frac{E}{\sigma}\right)^{\frac{3}{2}}-\frac{\pi}{4}\right] A_{b_1b_2},\,\,\,\,\,\,\,{\rm for}\,\,x>0,\\
\,\\
C\frac{1}{\left(x+\frac{E}{\sigma}\right)^{\frac{1}{4}}}\cos \left[-\frac{2}{3}(m\sigma)^{\frac{1}{2}}
\left(x+\frac{E}{\sigma}\right)^{\frac{3}{2}}+\frac{\pi}{4}\right]A_{b_1b_2},\,\,\,\,\,\,\,{\rm for}\,\,x<0.\end{array}\right.\label{wkbapprox}
\eeq
 By comparing (\ref{singletstate}) and (\ref{wkbapprox}) as $x\downarrow 0$, we fix
\beq
C=\left(\frac{E}{\sigma}\right)^{\frac{1}{4}},\,\,\,\,\,\,\,\omega=\frac{2}{3}(m\sigma)^{\frac{1}{2}}\left(\frac{E}{\sigma}\right)^{\frac{3}{2}}-\frac{\pi}{4}.\nonumber
\eeq
Comparing (\ref{singletstate}) and (\ref{wkbapprox}) as $x\uparrow 0$, gives the quantization condition.
\beq
\frac{4}{3}(m\sigma)^{\frac{1}{2}}\left(\frac{E}{\sigma}\right)^{\frac{3}{2}}+\frac{h_N}{\pi m}(m E)^{\frac{1}{2}}-\left(n+\frac{1}{4}\right)2\pi=0,\label{quantizationsinglet}
\eeq
with $n=0,1,2,...$. 

The solution to (\ref{quantizationsinglet}) is
\beq
E_n=\left\{\left[\epsilon_n+\left(\epsilon_n^2+\beta_N^3\right)^{\frac{1}{2}}\right]^{\frac{1}{3}}+\left[\epsilon_n-\left(\epsilon_n^2+\beta_N^3\right)^{\frac{1}{2}}\right]^{\frac{1}{3}}\right\}^{\frac{1}{2}},\label{spectrum1} \eeq
where
\beq
\epsilon_n=\frac{3\pi}{4}\left(\frac{\sigma}{m}\right)^{\frac{1}{2}}\left(n+\frac{1}{4}\right),\,\,\,\,\,\,\,\,\,\,\,\,\,
\beta_N=\frac{h_N\sigma^{\frac{1}{2}}}{4\pi m}. \label{spectrum2}
\eeq

\section{Horizontal Correlation Functions}
\setcounter{equation}{0}
\renewcommand{\theequation}{6.\arabic{equation}}

In this section we compute the long-distance correlation function of two gauge-invariant operators separated in the $x^1$-direction only. This is
\beq
\mathcal{D}^{\mathcal{A}}(x^1)=\langle 0\vert \mathcal{A}(x^1,x^2)\mathcal{A}(0,x^2)\vert0\rangle.\label{horizontalcorrelator}
\eeq
The correlation function (\ref{horizontalcorrelator}) can be evaluated by inserting a complete set of physical states between the two operators:
\beq
\mathcal{D}^{\mathcal{A}}(x^1)=\sum_\Psi \langle 0\vert \mathcal{A}(x^1,x^2)\vert \Psi\rangle\langle \Psi\vert\mathcal{A}(0,x^2)\vert0\rangle.\nonumber
\eeq

The physical, gauge invariant excitations of the theory are glueball bound states of sigma-model particles.  At large separations ($x^1\to\infty$), the function $\mathcal{D}^{\mathcal{A}}(x^1)$ can be approximated by inserting only one-glueball states. The lightest glue balls are those composed of a sigma-model particle and antiparticle, whose masses where calculated in the previous section. 

We denote the state with one glueball with rapidity $\phi$, and rest energy $M_n$, by $\vert B,\phi,n\rangle.$ The long-distance correlation function is
\beq
\mathcal{D}^{\mathcal{A}}(x^1)=\sum_{n=1}^{n_s}\int \frac{d\phi}{4\pi}\langle 0\vert\mathcal{A}(x^1,x^2)\vert B,\phi,n\rangle\langle B,\phi,n\vert \mathcal{A}(0,x^2)\vert 0\rangle,
\nonumber
\eeq
where $n_s$ is the energy level of the heaviest stable glueball, defined by $M_{n_s}\le2m\le M_{n_s+1}$.

We need a way to compute the one-glueball form factor of the operator $\mathcal{A}$. One approach was proposed by Fonseca and Zamolodchikov \cite{fonseca} in the Ising model perturbed by a weak external magnetic field. In the nonrelativistic limit, the glueball state is given by the so-called two-quark approximation:
\beq
\vert B, 0, n\rangle=\frac{1}{\sqrt{m}}\int_{-\infty}^\infty\frac{d\theta}{4\pi} \Psi_n(\theta)\,\vert A,\theta,a_1,b_1;P,-\theta,a_1,b_1\rangle,\nonumber
\eeq
where $\Psi_n(\theta)$ is the Fourier transform of the glueball wave function calculated in last section:
\beq
\Psi_n(\theta)=\int dz e^{izm\sinh\theta}\left(\frac{E_n}{\sigma^H}\right)^{\frac{1}{4}} {\rm Ai}\left[(m\sigma^H)^{\frac{1}{3}}\left(\vert z\vert -\frac{E_n}{\sigma^H}\right)\right].\nonumber
\eeq

If the operator $\mathcal{A}$ has spin $s$, the one-glueball form factor is
\beq
\langle 0\vert \mathcal{A}(x^1,x^2)\vert B,\phi,n\rangle&=&e^{s\phi}e^{ix^1M_n \sinh\phi}\int dz\int\frac{d\theta}{4\pi} e^{izm\sinh\theta}\frac{1}{\sqrt{m}}\left(\frac{E_n}{\sigma^H}\right)^{\frac{1}{4}} {\rm Ai}\left[(m\sigma^H)^{\frac{1}{3}}\left(\vert z\vert -\frac{E_n}{\sigma^H}\right)\right]\nonumber\\
&&\times\langle 0\vert \mathcal{A}(0,x^2)\vert A,\theta,a_1,b_1;P,-\theta,a_1,b_1\rangle.\nonumber
\eeq
For the rest of this section we will assume $s=0$.

The two-point correlation function is then
\beq
\mathcal{D}^{\mathcal{A}}(x^1)&=&\langle 0\vert \mathcal{A}(x^1,x^2)\vert 0\rangle \langle 0\vert \mathcal{A}(0,x^2)\vert 0\rangle\nonumber\\
&&+\sum_{n=1}^{n_s}\int\frac{d\phi}{4\pi}e^{ix^1M_n\sinh \phi}\left|\int dz\int\frac{d\theta}{4\pi} e^{izm\sinh\theta}\frac{1}{\sqrt{m}}\left(\frac{E_n}{\sigma^H}\right)^{\frac{1}{4}} {\rm Ai}\left[(m\sigma^H)^{\frac{1}{3}}\left(\vert z\vert -\frac{E_n}{\sigma^H}\right)\right]\mathcal{F}(2\theta)\right|^2,\nonumber
\eeq
where
\beq
\mathcal{F}(\theta)= C_{\mathcal{A}}N^2\frac{1}{(\theta+\pi i)}\exp\int_0^\infty\frac{d\xi}{\xi}\left[\frac{-2\sinh\left(\frac{2\xi}{N}\right)}{\sinh\xi}+\frac{4e^{-\xi}\left(e^{2\xi/N}-1\right)}{1-e^{-2\xi}}\right]\frac{\sin^2[\xi(\pi i-\theta)/2\pi]}{\sinh\xi},\nonumber
\eeq
and $C_\mathcal{A}$ is a normalization constant for the form factor of the operator $\mathcal{A}$.

The integral over $\phi$ gives
\beq
\int\frac{d\phi}{4\pi}e^{ix^1M_n\sinh\phi}=\frac{1}{2\pi}K_0(M_nx^1),\label{bessel}
\eeq
where $K_\alpha(x)$ is the modified Bessel function of the second kind. For $x^1\to\infty$, this modified Bessel function is approximated by 
\beq
K_0(M_n x^1)\to\sqrt{\frac{\pi}{2 M_n x^1}}e^{-M_n x^1}.\nonumber
\eeq

In the anisotropic limit where $\lambda\to0$, the glueball masses become similar to each other ($M_n\approx M_{n+1}$). Following \cite{bhaseen}, we assume that the form factors for any $n$ give approximately the same contribution. All the $n$ dependence is contained in the Bessel function (\ref{bessel}). In this limit, the sum over $n$ is approximated by a continuous integral, so we evaluate
\beq
\sum_{n=1}^{n_x}\sqrt{\frac{1}{M_nx^1}}e^{-M_nx}=\sum_{n=1}^{n_s}\sqrt{\frac{1}{2mx^1+E_nx^1}}e^{-2mx^1}e^{-E_nx^1}\approx\frac{e^{-2mx^1}}{\sqrt{2mx^1}}\int_0^\infty dn\, e^{-E_nx^1}\equiv\frac{e^{-2mx^1}}{\sqrt{2mx^1}}\mathcal{I}(x^1,\sigma^H,m).\label{continuousintegral}
\eeq
The integral $\mathcal{I}(x^1,\sigma^H,m)$ from Eq. (\ref{continuousintegral}) is in general quite complicated. One particularly simple case is when $N\to\infty$, where
\beq
\mathcal{I}(x^1,\sigma^H,m)=\int_0^\infty dn\,\exp\left\{\left[\frac{3\pi}{4}\left(\frac{\sigma^H}{m}\right)^{\frac{1}{2}}n\right]^\frac{1}{6}\right\}=\frac{720}{\frac{3\pi}{4}\left(\frac{\sigma^H}{m}\right)^{\frac{1}{2}}x^6}.\nonumber
\eeq

The horizontal correlation function at large distances and small $\lambda$ is 
\beq
\mathcal{D}^{\mathcal{A}}(x^1)=&=&\langle 0\vert \mathcal{A}(x^1,x^2)\vert 0\rangle \langle 0\vert \mathcal{A}(0,x^2)\vert 0\rangle\nonumber\\
&&+\frac{e^{-2mx^1}}{\sqrt{8m^5x^1}}\mathcal{I}(x^1,\sigma^H,m)\left(\frac{E_0}{\sigma^H}\right)^{\frac{1}{2}}\left|\int dz\int\frac{d\theta}{4\pi} e^{izm\sinh\theta}{\rm Ai}\left[(m\sigma^H)^{\frac{1}{3}}\left(\vert z\vert -\frac{E_n}{\sigma^H}\right)\right]\mathcal{F}(2\theta)\right|^2.\nonumber
\eeq

\section{Vertical Correlation Functions}
In this section we present a method to evaluate correlation functions of two gauge-invariant operators separated in the $x^2$-direction only. This problem is significantly harder than calculating horizontal correlation functions, and we are able to make progress only in the large-$N$ limit. For large separation in $x^2$, the problem is reduced to solving an integral eigenvalue equation.

We want  to calculate the correlator
\beq
\mathcal{D}^{\mathcal{A}}(aR^2)=\langle 0\vert \mathcal{A}(x^1,x^2)\mathcal{A}(x^1,x^2+aR^2)\vert 0\rangle.\label{correlationxtwo}
\eeq
Our strategy is to define a transfer matrix operator, $T_{x^2,x^2+a}$, that describes the evolution of the system in the $x^2$-direction. We impose periodic boundary conditions in the $x^2$-direction. We call size of the $x^2$-dimension $L_2$. The partition function and correlation functions can be computed by diagonalizing this transfer matrix.

The transfer matrix is defined by
\beq
T_{x^2,x^2+a}=e^{-H_{x^2,x^2+a}},\nonumber
\eeq
where
\beq
H_{x^2,x^2+a}&=&\frac{1}{2}H_{\rm PCSM}(x^2)+\frac{1}{2}H_{\rm PCSM}(x^2+a)+\lambda^2H_1(x^2,x^2+a)\nonumber
\eeq
and 
\beq
H_1(x^2,x^2+a)&=&\int dx^1 dy^1 \frac{1}{8g_0^2a}\vert x^1-y^1\vert \left[j_0^L(x^1,x^2)j_0^L(y^1,x^2)+j_0^R(x^1,x^2)j_0^R(y^1,x^2)\right]\nonumber\\
&&+\int dx^1 dy^1 \frac{1}{8g_0^2a}\vert x^1-y^1\vert\left[j_0^L(x^1,x^2+a)j_0^L(y^1,x^2+a)+j_0^R(x^1,x^2+a)j_0^R(y^1,x^2+a)\right]\nonumber\\
&&-\int dx^1 dy^1 \frac{1}{4g_0^2a}\vert x^1-y^1\vert \left[j_0^L(x^1,x^2+a)j_0^R(y^1,x^2)+j_0^L(y^1,x^2+a)j_0^R(x^1,x^2)\right].\label{interactiontransfer}
\eeq

We can now compute the matrix elements in of $T_{x^2,x^2+a}$ between the particle states of the sigma models at $x^2$ and $x^2+a$. We label a state with a two-particle bound state, of rapidity $\phi$ and energy level $n$ on the sigma model at $x^2$ by
$\vert B,\phi,n,x^2\rangle$. For large separations in $x^2$, it is sufficient to compute the matrix elements with just one bound state in each $x^2$-layer.

We define the functions
\beq
T&=&\langle 0\vert T_{x^2,x^2+a}\vert 0\rangle,\nonumber\\
T_n(\phi)&=&\langle B,\phi,n,x^2\vert T_{x^2,x^2+a}\vert B,\phi,n,x^2\rangle,\nonumber\\
T_{nn^\prime}(\phi,\phi^\prime)&=&\langle B,\phi,n,x^2;B,\phi^\prime,n^\prime,x^2+a\vert T_{x^2,x^2+a}\vert B,\phi,n,x^2;B,\phi^\prime,n^\prime,x^2+a\rangle.\label{definitionts}
\eeq
In the basis of one-glueball states, the transfer matrix is
\beq
\tau_{nn^\prime}(\phi,\phi^\prime)=T+T_n(\phi)+T_{n^\prime}(\phi^\prime)+T_{nn^\prime}(\phi,\phi^\prime).\label{transfermatrix}
\eeq
The partition function and correlation functions can be found, in principle, by finding the eigenvalues of the matrix $\tau_{nn^\prime}(\phi,\phi^\prime)$. This means one has to solve the integral equation
\beq
\sum_{n^\prime=1}^{n_s}\int\frac{d\phi^\prime}{4\pi}\tau_{nn^\prime}(\phi,\phi^\prime)\psi_{n^\prime}^{(l)}(\phi^\prime)=\lambda^{(l)}\psi_{n}^{(l)}(\phi).\label{eigenvalueequation}
\eeq

If the eigenvalues $\lambda^{(l)}$ and eigenfunctions $\psi_n^{(l)}(\phi)$ are known, the transfer matrix may be diagonalized as
\beq
\tau_{nn^\prime}(\phi,\phi^\prime)=\sum_l\lambda^{(l)}\psi_n^{(l)}(\phi)\psi_{n^\prime}^{(l)}(\phi^\prime),\label{diagonalizedtransfermatrix} 
\eeq
where the eigenfunctions are normalized by
\beq
\sum_n\int\frac{d\phi}{4\pi}\psi_n^{(l)}(\phi)\psi_n^{(m)}(\phi)=\delta^{lm}.\nonumber
\eeq
The partition function is then given by 
\beq
Z=\sum_l\left[\lambda^{(l)}\right]^{N^2},\nonumber
\eeq
where 
\beq
N^2=\frac{L_2}{a}.\nonumber
\eeq
In the thermodynamic limit, $N^2\to\infty$, the partition function is 
\beq
Z=[\lambda^{(0)}]^{N^2},\nonumber
\eeq
where $\lambda^{(0)}$ is the largest eigenvalue of the transfer matrix.

The operators in the correlation function (\ref{correlationxtwo}) are expressed in the one-glueball basis as the functions $\mathcal{A}_{nn^\prime}(\phi,\phi^\prime)$. We assume that the functions $\mathcal{A}$ and $\tau$ are not simultaneously diagonalizable. The two-point correlation function is
\beq
\mathcal{D}^{\mathcal{A}}(aR^2)&=&\frac{1}{\left[\lambda^{(0)}\right]^{N^2}}\sum_{l,l_1,l_2}\,\,\,\,\sum_{n_1,n_2,n_{R^2},n_{R^2+1}}\int\frac{d\phi_1 d\phi_2 d\phi_{R^2} d\phi_{R^2+1}}{(4\pi)^4}\nonumber\\
&&\times\left\{\psi_{n_1}^{(l_1)}(\phi_1)\mathcal{A}_{n_1n_2}(\phi_1,\phi_2)\psi_{n_2}^{(l_2)}(\phi_2)[\lambda^{(l)}]^{R^2-3}\psi_{n_{R^2}}^{(l_2)}(\phi_{R^2})\mathcal{A}_{n_{R^2}\,n_{R^2+1}}(\phi_{R^2},\phi_{R^2+1})\psi_{n_{R^2+1}}^{(l_1)}(\phi_{R^2+1})\right\}.\label{xtwocorr}
\eeq
In the limit of large separation $R^2$, Eq. (\ref{xtwocorr}) becomes
\beq
\mathcal{D}^{\mathcal{A}}(aR^2)=\mathcal{C}+\left(\frac{\lambda^{(1)}}{\lambda^{(0)}}\right)^{R^2}\mathcal{C},\label{largertwolimit}
\eeq
where $\lambda^{(1)}$ is the second largest eigenvalue, and 
\beq
\mathcal{C}&=&\sum_{l_1, l_2}\,\,\,\,\sum_{n_1,n_2,n_{R^2},n_{R^2+1}}\int\frac{d\phi_1 d\phi_2 d\phi_{R^2} d\phi_{R^2+1}}{(4\pi)^4}\nonumber\\
&&\times\left\{\psi_{n_1}^{(l_1)}(\phi_1)\mathcal{A}_{n_1n_2}(\phi_1,\phi_2)\psi_{n_2}^{(l_2)}(\phi_2)\psi_{n_{R^2}}^{(l_2)}(\phi_{R^2})\mathcal{A}_{n_{R^2}\,n_{R^2+1}}(\phi_{R^2},\phi_{R^2+1})\psi_{n_{R^2+1}}^{(l_1)}(\phi_{R^2+1})\right\}.\nonumber
\eeq
We define the inverse correlation length $\mathcal{M}$ as
\beq
\mathcal{D}^{\mathcal{A}}(aR^2)\sim e^{-\mathcal{M}aR^2}.\nonumber
\eeq
From Eq. (\ref{largertwolimit}), the inverse correlation length is
\beq
\mathcal{M}=-\frac{1}{a}\ln\left(\frac{\lambda^{(1)}}{\lambda^{(0)}}\right).\nonumber
\eeq

The rest of this section is dedicated to finding an expression for $\tau_{nn^\prime}(\phi,\phi^\prime)$, though we are never able to solve the integral eigenvalue equation (\ref{eigenvalueequation}). This is left as an open problem.

The contribution to $T_{nn^\prime}(\phi,\phi^\prime)$ which couples two adjacent $x^2$ layers involves the two-bound state form factor of the current operator. This means that we need the four-excitation form factors of the PCSM. The functions $T,\,T_n(\phi),\,T_{nn^\prime}(\phi,\phi^\prime)$ involve two-point functions of current operators. These correlation functions will be computed keeping only terms proportional to the two-and four-particle form factors. Form factors of more than two excitations are only known in  't Hooft's large-$N$ limit. For the rest of this section we work exclusively in the large-$N$ limit. The form factors of the sigma model at large $N$ were found in references \cite{multiparticle}, \cite{myselforland}, \cite{renormalizedfield}, and are reviewed in the appendix.

We first calculate the constant 
\beq
T&=&\langle 0\vert e^{-\frac{1}{2}H_{\rm PCSM}(x^2)-\frac{1}{2}H_{\rm PCSM}(x^2+a)+\lambda^2H_1(x^2,x^2+a)}\vert 0\rangle\nonumber\\
&\approx& \exp\left\{\langle 0\vert -\frac{1}{2}H_{\rm PCSM}(x^2)-\frac{1}{2}H_{\rm PCSM}(x^2+a)-\lambda^2H_1(x^2,x^2+a)\vert 0\rangle\right\}.\label{tfunction}
\eeq
The constant $T$ only has a contribution from $H_1(x^2,x^2+a)$. This contribution is
\beq
\langle 0\vert H_1(x^2,x^2+a)\vert 0\rangle&=&\int dx^1 dy^1\vert x^1-y^1\vert\frac{1}{4g_0^2a}\left\{\langle 0\vert j_0^L(x^1,x^2)j_0^L(y^1,x^2)\vert0\rangle+\langle 0\vert j_0^R(x^1,x^2)j_0^R(y^1,x^2)\vert0\rangle\right\}\nonumber\\
&&\!\!\!\!\!\!\!\!\!\!\!\!\!\!\!\!\!\!\!\!\!\!\!\!\!\!\!\!\!\!\!\!\!\!\!\!\!\!\!+\int dx^1 dy^1\vert x^1-y^1\vert\frac{1}{4g_0^2a}\left\{\langle 0\vert j_0^L(x^1,x^2+a)j_0^L(y^1,x^2+a)\vert0\rangle+\langle 0\vert j_0^R(x^1,x^2+a)j_0^R(y^1,x^2+a)\vert0\rangle\right\}.\label{t}
\eeq

We now examine the correlation functions in the right hand side of Eq. (\ref{t}) using up to two-glueball form factors. This is, for the left-handed current,
\beq
\langle 0\vert j_0^L(x^1,x^2)j_0^L(y^1,x^2)\vert 0\rangle&=&\sum_n\int\frac{d\phi}{4\pi}\langle 0\vert j_0^L(x^1,x^2)\vert B,\phi,n,x^2\rangle\langle B,\phi, n,x^2\vert j_0^L(y^1,x^2)\vert 0\rangle\nonumber\\
&&+\sum_{n_1,n_2}\int\frac{d\phi_1\,d\phi_2}{(4\pi)^2}\left[\langle 0\vert j_0^L(x^1,x^2)\vert B,\phi_1,n_1,x^2;B,\phi_2,n_2,x^2\rangle\right.\nonumber\\
&&\times\left.\langle B,\phi_1,n_1,x^2;B,\phi_2,n_2,x^2\vert j_0^L(y^1,x^2)\vert 0\rangle\right].\nonumber
\eeq

Using the large-N, two-excitation form factors of the sigma model (found in the appendix), we find
\beq
&&\sum_n\int\frac{d\phi}{4\pi}\langle 0\vert j_0^L(x^1,x^2)\vert B,\phi,n,x^2\rangle\langle B,\phi, n,x^2\vert j_0^L(y^1,x^2)\vert 0\rangle\nonumber\\
&&\,\,\,\,\,\,=\sum_n\int\frac{d\phi}{4\pi} M_n\sinh^2\phi e^{-i(x^1-y^1)M_n\sinh\phi}\left(\frac{E_n}{\sigma}\right)^{\frac{1}{4}}N^2
\left|\int dz\int\frac{d\theta}{4\pi}{\rm Ai}\left[(m\sigma)^{\frac{1}{3}}\left(\vert z\vert-\frac{E_n}{\sigma}\right)\right]\frac{2\pi i \tanh\theta}{2\theta+\pi i}\right|^2\nonumber.
\eeq
The contribution to $T$ from the one-glueball form factor is 
\beq
T^{(2)}&=&\exp\left\{-4\lambda^2\int dx^1 dy^1\vert x^1-y^1\vert\frac{1}{4g_0^2 a}\sum_n\int\frac{d\phi}{4\pi}M_n\sinh^2\phi e^{-i(x^1-y^1)M_n\sinh\phi}\left(\frac{E_n}{\sigma}\right)^{\frac{1}{4}}N^2\mathcal{F}^{(2)}\right\},\label{resultttwo}
\eeq
where
\beq
\mathcal{F}^{(2)}=\left|\int dz\int\frac{d\theta}{4\pi}{\rm Ai}\left[(m\sigma)^{\frac{1}{3}}\left(\vert z\vert-\frac{E_n}{\sigma}\right)\right]\frac{2\pi i \tanh\theta}{2\theta+\pi i}e^{izm\sinh\theta}\right|^2.\nonumber
\eeq

Using the large-N, four-excitation form factors of the sigma model, we find
\beq
&&\sum_{n_1,n_2}\int\frac{d\phi_1\,d\phi_2}{(4\pi)^2}\langle 0\vert j_0^L(x^1,x^2)\vert B,\phi_1,n_1,x^2;B,\phi_2,n_2,x^2\rangle\langle B,\phi_1,n_1,x^2;B,\phi_2,n_2,x^2\vert j_0^L(y^1,x^2)\vert 0\rangle\nonumber\\
&&=\sum_{n_1,n_2}\int\frac{d\phi_1\,d\phi_2}{(4\pi)^2}\left(\frac{E_{n_1}E_{n_2}}{\sigma^2}\right)^{\frac{1}{2}}\frac{1}{\sqrt{M_{n_1}M_{n_2}}}\left(M_{n_1}\sinh\phi_1+M_{n_2}\sinh\phi_2\right)^2e^{-i(x^1-y^1)(M_{n_1}\sinh\phi_1+M_{n_2}\sinh\phi_2)}N^2\mathcal{F}^{(4)},\nonumber
\eeq
where
\beq
\mathcal{F}^{(4)}&=&\left|\int \frac{d\theta d\theta^\prime}{(4\pi)^2}\frac{\tanh\theta f(\theta,\theta^\prime)}{\left[(\theta+\theta^\prime)^2+\pi^2\right](2\theta^\prime+\pi i)}\right|^2+\left|\int \frac{d\theta d\theta^\prime}{(4\pi)^2}\frac{\tanh\left(\frac{\theta+\theta^\prime}{2}\right)f(\theta,\theta^\prime)}{(2\theta+\pi i)(\theta^\prime+\theta-\pi i)(2\theta^\prime+\pi i)}\right|^2\nonumber\\
&&+\left|\int \frac{d\theta d\theta^\prime}{(4\pi)^2}\frac{\tanh\left(\frac{\theta+\theta^\prime}{2}\right)f(\theta,\theta^\prime)}{(\theta+\theta^\prime+\pi i)(2\theta^\prime+\pi i)(2\theta+\pi i)}\right|^2+\left|\int \frac{d\theta d\theta^\prime}{(4\pi)^2}\frac{\tanh\theta^\prime f(\theta,\theta^\prime)}{\left[(\theta+\theta^\prime)^2+\pi^2\right](2\theta+\pi i)}\right|^2,\nonumber
\eeq
and
\beq
f(\theta,\theta^\prime)=\int dz_1 dz_2 \,8\pi i\,e^{iz_1 m\sinh\theta+iz_2\sinh\theta^\prime}{\rm Ai}\left[(m\sigma)^{\frac{1}{3}}\left(\vert z_1\vert-\frac{E_n}{\sigma}\right)\right]{\rm Ai}\left[(m\sigma)^{\frac{1}{3}}\left(\vert z_2\vert-\frac{E_n}{\sigma}\right)\right].\nonumber
\eeq
The contribution to $T$ from the two-glueball form factor is
\beq
T^{(4)}&=&\exp\left\{-4\lambda^2\int dx^1dy^1\vert x^1-y^1\vert\frac{1}{4g_0^2 a}\sum_{n_1,n_2}\int\frac{d\phi_1\,d\phi_2}{(4\pi)^2}\left(\frac{E_{n_1}E_{n_2}}{\sigma^2}\right)^{\frac{1}{2}}\right.\nonumber\\
&&\times\left.\frac{1}{\sqrt{M_{n_1}M_{n_2}}}\left(M_{n_1}\sinh\phi_1+M_{n_2}\sinh\phi_2\right)^2e^{-i(x^1-y^1)(M_{n_1}\sinh\phi_1+M_{n_2}\sinh\phi_2)}N^2\mathcal{F}^{(4)}\right\},\label{resulttfour}
\eeq
such that
\beq
T=T^{(2)}T^{(4)}.\label{resultt}
\eeq

We now calculate the function $T_n(\phi)$ from Eq. (\ref{definitionts}). The contribution to $T_n(\phi)$ from the sigma model in the $x^2+a$ layer is just $\sqrt{T^{(2)}T^{(4)}}$. There is a contribution to the function $T_n(\phi)$ from the unperturbed Hamiltonian, given by 
\beq
&&\langle B,\phi,n,x^2\vert H_{\rm PCSM}(x^2)\vert B,\phi,n,x^2\rangle\equiv T_n^{(0)}(\phi)\nonumber\\
&&=\frac{1}{M_n}\int dz_1dz_2\left(\frac{E_N}{\sigma}\right)^{\frac{1}{2}}{\rm Ai}\left[(m\sigma)^{\frac{1}{3}}\left(\vert z_1\vert -\frac{E_n}{\sigma}\right)\right]{\rm Ai}\left[(m\sigma)^{\frac{1}{3}}\left(\vert z_2\vert-\frac{E_n}{\sigma}\right)\right]\nonumber\\
&&\,\,\,\,\,\,\,\times\int\frac{d\theta}{4\pi}2m\cosh\theta e^{i(z_2-z_1)m\sinh\theta}e^{i(z_1-z_2)M_n\cosh\phi}.\label{resulttnzero}
\eeq

There are contributions to $T_n(\phi)$ from the current correlation function
\beq
&&\langle B,\phi,n,x^2\vert j_0^L(x^1,x^2)j_0^L(y^1,x^2)\vert B,\phi,n,x^2\rangle\nonumber\\
&&=\langle B,\phi,n,x^2\vert j_0^L(x^1,x^2)\vert0\rangle\langle0\vert j_0^L(y^1,x^2)\vert B,\phi,n,x^2\rangle\nonumber\\
&&+\sum_n^\prime\int \frac{d\phi^\prime}{4\pi}\langle B,\phi,n,x^2\vert j_0^L(x^1,x^2)\vert B,\phi^\prime,n^\prime,x^2\rangle\langle B,\phi^\prime,n^\prime,x^2\vert j_0^L(y^1,x^2)\vert B,\phi,n,x^2\rangle+\dots\,\,.\nonumber
\eeq
Using the two-excitation form factors of the sigma model, we find
\beq
\langle B,\phi,n,x^2\vert j_0^L(x^1,x^2)\vert0\rangle\langle0\vert j_0^L(y^1,x^2)\vert B,\phi,n,x^2\rangle=M_n\sinh^2\phi\,e^{-i(x^1-y^1)M_n\sinh\phi}\left(\frac{E_n}{\sigma}\right)^{\frac{1}{4}}\mathcal{F}^{(2)}.\nonumber
\eeq
The contribution to $T_n(\phi)$ from the one-glueball form factors of the sigma model at $x^2$ is
\beq
T_n^{(2)}(\phi)=\exp\left\{-2\lambda^2\int dx^1dy^1\vert x^1-y^1\vert\frac{1}{4g_0^2a}M_n\sinh^2\phi\,e^{-i(x^1-y^1)M_n\sinh\phi}\left(\frac{E_n}{\sigma}\right)^{\frac{1}{4}}\mathcal{F}^{(2)}\right\}.\label{resulttntwo}
\eeq

Using the form factor of the sigma model with two incoming and two outgoing excitations (also found in the appendix), we find the contribution $T_n(\phi)$ from the two-glueball form factor
\beq
&&T_n^{(4)}(\phi)=\exp\left\{-2\lambda^2\int dx^1 dy^1\vert x^1-y^1\vert\frac{1}{4g_0^2 a}\sum_{n^\prime}\int\frac{d\phi}{4\pi}\left(\frac{E_nE_{n^\prime}}{\sigma^2}\right)^{\frac{1}{2}}N^2\right.\nonumber\\
&&\,\,\,\,\,\,\,\,\,\,\times \left.\frac{1}{\sqrt{M_n M_{n^\prime}}}\left(M_n\sinh\phi+M_{n^\prime}\sinh\phi^\prime\right)^2\,e^{-i(x^1-y^1)(-M_n\sinh\phi+M_{n^\prime}\sinh\phi^\prime)}\mathcal{F}^{\prime\,(4)}\right\},\label{resulttnfour}
\eeq
where
\beq
\mathcal{F}^{\prime\,(4)}&=&\left|\int\frac{d\theta d\theta^\prime}{(4\pi)^2}\frac{\tanh\theta f^\prime(\theta,\theta^\prime)}{(\theta+\theta^\prime+2\pi i)(\theta^\prime+\theta-2\pi i)(2\theta^\prime+\pi i)}\right|^2+\left|\int\frac{d\theta d\theta^\prime}{(4\pi)^2}\frac{\coth\left(\frac{\theta+\theta^\prime}{2}\right) f^\prime(\theta,\theta^\prime)}{(2\theta+\pi i)(\theta^\prime+\theta-2\pi i)(2\theta^\prime+\pi i)}\right|^2\nonumber\\
&&+\left|\int\frac{d\theta d\theta^\prime}{(4\pi)^2}\frac{\coth\left(\frac{\theta+\theta^\prime}{2}\right) f^\prime(\theta,\theta^\prime)}{\theta+\theta^\prime+2\pi i)(2\theta+\pi i)(2\theta^\prime+\pi i)}\right|^2+\left|\int\frac{d\theta d\theta^\prime}{(4\pi)^2}\frac{\tanh\theta^\prime f^\prime(\theta,\theta^\prime)}{(\theta+\theta^\prime+2\pi i)(2\theta+\pi i)(\theta+\theta^\prime-2\pi i)}\right|^2,\nonumber
\eeq
and
\beq
f^\prime(\theta,\theta^\prime)=8\pi i\int dz_1 {\rm Ai}\left[(m\sigma)^{\frac{1}{3}}\left(\vert z_1\vert-\frac{E_n}{\sigma}\right)\right]e^{iz_1 m\sinh\theta}\left\{\int dz_2 {\rm Ai}\left[(m\sigma)^{\frac{1}{3}}\left(\vert z_2\vert-\frac{E_n}{\sigma}\right)\right]e^{iz_2 m\sinh\theta^\prime}\right\}^*.\nonumber
\eeq
We can combine the results from Equations (\ref{resultt}), (\ref{resulttnzero}), (\ref{resulttntwo}), and (\ref{resulttnfour}) to write
\beq
T_n(\phi)=\sqrt{T}\,T_n^{(0)}(\phi)T_n^{(2)}(\phi)T_n^{(4)}(\phi).\label{resulttn}
\eeq

We now evaluate the function $T_{nn^\prime}(\phi,\phi^\prime)$. This function has only one new contribution, which couples between the $x^2$ and $x^2+a$ layers. This is
\beq
K_{nn^\prime}(\phi,\phi^\prime)&=&\exp\left\{-\lambda^2\int dx^1dy^1\vert x^1-y^1\vert \frac{1}{2g_0^2 a}\langle B, \phi,n,x^2\vert j_0^R(x^1,x^2)\vert B,\phi,n,x^2\rangle\right.\nonumber\\
&&\times\left.\langle B,\phi^\prime,n^\prime,x^2+a\vert j_0^L(y^1,x^2+a)\vert B,\phi^\prime,n^\prime,x^2+a\rangle\right\}\nonumber\\
&=&\exp\left\{-\lambda^2\int dx^1dy^1\vert x^1-y^1\vert \frac{1}{2g_0^2 a}\left(\frac{E_nE_{n^\prime}}{\sigma^2}\right)^{\frac{1}{2}}N^2\right.\nonumber\\
&&\times\left.M_nM_{n^\prime}\,\sinh^2\phi \sinh^2\phi^\prime\,e^{-ix^1M_n\sinh\phi+iy^1M_{n^\prime}\sinh\phi^\prime}\mathcal{F}^{\prime\,(4)}\right\}.\label{resultk}
\eeq
With Eq. (\ref{resultk}) and  Equations (\ref{resultt}), (\ref{resulttnzero}), (\ref{resulttntwo}), and (\ref{resulttnfour}), we write
\beq
T_{nn^\prime}(\theta,\theta^\prime)=T_n^{(0)}(\phi)T_n^{(2)}(\phi)T_n^{(4)}(\phi)\,T_{n^\prime}^{(0)}(\phi^\prime)T_{n^\prime}^{(2)}(\phi^\prime)T_{n^\prime}^{(4)}(\phi^\prime)\,K_{nn^\prime}(\phi,\phi^\prime).\label{resulttnn}
\eeq

The transfer matrix $\tau_{nn^\prime}(\phi,\phi)$ is given in Eq. (\ref{transfermatrix}), by combining Equations (\ref{resultt}), (\ref{resulttn}), (\ref{resulttnn}). The problem of finding the vertical correlation functions is now reduced to diagonalizing the function $\tau_{nn^\prime}(\phi,\phi)$, and expressing it in the form of Eq. (\ref{diagonalizedtransfermatrix}).

\section{Conclusion}

We have used new exact results from the principal chiral sigma model to compute physical quantities in anisotropic Yang-Mills theory. The two-particle form factors of the sigma model are now known for general $N>2$. This allowed us to generalize Orland's results for the $SU(2)$ gauge group, to $SU(N)$.  These results include the string tensions for quarks separated in the $x^1$ and $x^2$-directions, and the spectrum of the lightest glueball masses.

Once we found the glueball states, we used them to calculate correlation functions of gauge invariant operators. For two operators separated in the $x^1$-direction only, the correlation function is calculated at long distances by summing over a complete set of intermediate one-glueball states.

The correlation functions of operators separated in the $x^2$-direction are much harder to calculate. We proposed a method for how these correlation functions may be calculated, though we do not solve the problem completely. We compute the elements of a transfer matrix which evolves the system in the $x^2$-direction.  These elements are computed in the basis of one-glueball states. The problem of calculating correlation functions is reduced to solving an integral eigenvalue equation for the transfer matrix.

An obvious problem for the future is to find a solution to the eigenvalue equation , Eq. (\ref{eigenvalueequation}). This would allow us to calculate explicitly the partition function and correlation functions in the $x^2$-direction. The rapidities of the glueballs, $\phi,\,\phi^\prime$ can be discretized by placing the sigma models in a finite box of size $L_1$. One can impose an energy cutoff by discarding glueball states above some maximum rapidity. The transfer matrix then becomes discrete and finite, and can thus be diagonalized numerically on a computer. This computation would be similar to that done for the Ising model by Konik and Adamov \cite{konikadamov}.
We would like to point out that the methods of Ref. \cite{konikadamov} can, in principle, be used to find results applicable to the fully isotropic (2+1)- dimensional theory. In this reference, the authors studied the three-dimensional Ising model as an array of two-dimensional chains, for different values of the interchain coupling (corresponding to our parameter $\lambda$), up to the fully isotropic value. Their transfer matrix was obtained by an improved version of the truncated spectrum approach \cite{TSA}. One difficulty for the Yang-Mills-theory case is that gauge invariance needs to be imposed on the states of the truncated spectrum, making the construction of the transfer matrix non trivial. This numerical diagonalization is the most promising approach that we know with which we could study the fully isotropic (2+1)-dimensional theory.

It would be interesting to extend our methods to 3+1 dimensions. It has been shown that longitudinally rescaled (3+1)-dimensional Yang-Mills theory can also be expressed as an array of sigma models \cite{threeplusone}. There is an additional interaction term given by the additional components of the magnetic field. It would be interesting to see what is the effect of this additional interaction on the quantities calculated in this paper.

\begin{acknowledgements}
I wish to thank Peter Orland for our many discussions about his earlier work that were so essential to this project, and for carefully reading this manuscript. I also thank Adrian Dumitru for some interesting discussions about the applications of anisotropic gauge theories. Lastly, I thank Robert Konik, for some very useful conversations we had over a year ago, which inspired some of this work.
\end{acknowledgements}

\appendix*
\section{The S-matrix and Form Factors of the Principal Chiral Sigma Model}
\setcounter{equation}{0}
\renewcommand{\theequation}{A.\arabic{equation}}
 For the purposes of this paper, we only need to know the two-and four-particle form factors of the Noether current operators of the sigma model. These were found in the 't Hooft limit in Ref. \cite{multiparticle}. For finite $N$, only the two-particle form factor is found in the same paper. These results were later generalized to form factors of an arbitrary number of particles, at large $N$, if Ref. \cite{myselforland}. These form factors were used to calculate two-point correlation functions. It is worth mentioning that the form factors and correlation functions of other operators have also been found in the 't Hooft limit. The renormalized field operator was studied in Reference \cite{renormalizedfield}, and the energy-momentum tensor was studied in \cite{myselforland}.

This appendix is not meant to be a review of form factors of integrable theories. We merely present results without a meticulous derivation. For a complete derivation of the results in this appendix, see Ref. \cite{multiparticle}. A  modern review of the integrable bootstrap program for calculating form factors are found in References \cite{babujian}.

The derivation of the form factors makes use of the two-particle S-matrix of the sigma model. This S-matrix has been found in Refs. \cite{wiegmann} \cite{integrablePCSM}.  The S-matrix, $S_{PP}(\theta)^{c_2d_2;c_1d_1}_{a_1b_1a_2b_2}$ of two incoming particles with rapidities $\theta_1$, and $\theta_2$ and left and right color indices $a_1,b_1$, and $a_2,b_2$ respectively, and two outgoing particles with rapidities $\theta_1^\prime$ and $\theta_2^\prime$, and left and right color indices $c_1,d_1$, and $c_2,d_2$, respectively,  is given by
\beq
\,_{\rm out}\langle P, \theta_1^\prime,c_1,d_1; A,\theta_2^\prime,d_2,c_2\!\!\!\!\!&\vert&\!\!\!\!\! P, \theta_1,a_1,b_1; A,\theta_2,b_2,a_2\rangle_{\rm in}\nonumber\\
&=&S_{PP}(\theta)_{a_1 b_1;a_2b_2}^{c_2d_2;c_1d_1}\,4\pi\,\delta(\theta_1^\prime-\theta_1)\,4\pi\,\delta(\theta_2^\prime-\theta_2),\nonumber
\eeq
where $\theta=\theta_1-\theta_2$. The result from \cite{wiegmann}, \cite{integrablePCSM} is 
\beq
S_{PP}(\theta)_{a_1 b_1;a_2b_2}^{c_2d_2;c_1d_1}=\chi(\theta) \,S_{\rm CGN}(\theta)_{a_1;a_2}^{c_2;c_1}\,S_{\rm CGN}(\theta)_{b_1;b_2}^{d_2d_1},\nonumber
\eeq
where $S_{\rm CGN}$ is the S-matrix of two elementary excitations of the $SU(N)$ chiral Gross-Neveu model \cite{berg}, \cite{kurak}:
\beq
S_{\rm CGN}(\theta)_{a_1a;a_2}^{c_2c_1}=\frac{\Gamma(i\theta/2\pi+1)\Gamma(-i\theta/2\pi-1/N)}{\Gamma(i\theta/2\pi+1-1/N)\Gamma(-i\theta/2\pi)}\left(\delta_{a_1}^{c_1}\delta_{a_2}^{c_2}-\frac{2\pi i}{N\theta}\delta_{a_2}^{c_1}\delta_{a_1}^{c_2}\right),\nonumber
\eeq
and 
\beq
\chi(\theta)=\frac{\sinh\left(\frac{\theta}{2}-\frac{\pi i}{N}\right)}{\sinh\left(\frac{\theta}{2}+\frac{\pi i}{N}\right)}. \nonumber
\eeq

The particle-antiparticle S-matrix is related to the particle-particle S-matrix by crossing symmetry, i.e. $\theta\to\hat{\theta}=\pi i-\theta$. It was found in Ref. \cite{multiparticle}, that the particle-antiparticle S-matrix can be written in the exponential form:
\beq
S(\theta)_{a_1b_1;b_2a_2}^{d_2c_2;c_1d_1}=S(\theta)\left[\delta_{a_1}^{c_1}\delta_{a_2}^{c_2}-\frac{2\pi {\rm i}}{N(\pi {\rm i}-\theta)}\delta_{a_1a_2}\delta^{c_1c_2}\right]\left[\delta_{b_1}^{d_1}\delta_{b_2}^{d_2}-\frac{2\pi {\rm i}}{N(\pi {\rm i}-\theta)}\delta_{b_1b_2}b^{d_1d_2}\right],\nonumber
\eeq
where
\beq
S(\theta)=\exp2 \int_0^\infty\,\frac{d\xi}{\xi \sinh\xi}\left[ 2(e^{2\xi/N}-1)-\sinh(2\xi/N)
\right]\sinh\frac{\xi \theta}{\pi{\rm i}} \;,
\label{Stheta}
\eeq
for $N>2$.

The two-particle form factor of the left-handed Noether current was found using Eq. (\ref{Stheta}). This is \cite{multiparticle}
\beq
\langle 0\vert  j_\mu^L(x)_{a_0c_0}\!\!\!\!\!&\vert&\!\!\!\!\! A,\theta_1,b_1,a_1;P,\theta_2,a_2,b_2\rangle_{\rm in}\nonumber\\
&=&(p_1-p_2)_\mu e^{-ix\cdot(p_1+p_2)}\left(\delta_{a_0a_2}\delta_{c_0a_1}\delta_{b_1b_2}-\frac{1}{N}\delta_{a_0c_0}\delta_{a_1a_2}\delta_{b_1b_2}\right).\nonumber\\
&&\times\frac{2\pi i}{(\theta+\pi i)}\exp\int_0^\infty \frac{dx}{x}\left[\frac{-2\sinh\left(\frac{2x}{N}\right)}{\sinh x}+\frac{4e^{-x}\left(e^{2x/N}-1\right)}{1-e^{-2x}}\right]\frac{\sin^2[x(\pi i-\theta)/2\pi]}{\sinh x}.\label{finitenform}
\eeq
The form factor with one incoming and one outgoing antiparticle can be found by crossing the particle in (\ref{finitenform}) into an outgoing particle, shifting the rapidity $\theta_2\to\theta_2-\pi i$. The right-handed current has a very similar expression, but the color indices of the operator are contracted with the right-handed color indices of the particle and antiparticle.

Next we show the four-excitation form factor at large $N$. The form factor is nonzero only if two of the excitations are particles ant two are antiparticles. The form factor is\footnote{It is important to mention that the four-particle form factor from Ref. \cite{multiparticle} has been found to not be completely correct as written. The momentum-vector prefactor chosen in \cite{multiparticle} is $(p_1+p_2-p_3-p_4)_\mu$ instead of $-\epsilon_{\mu\nu}(p_1+p_2+p_3+p_4)^\nu$, as we have written in Eq. (\ref{finalanswerfourparticle}). The results from References \cite{multiparticle} and \cite{myselforland} are not consistent with the fact that the Noether current is conserved. These corrections have been published in \cite{thesis}}
\beq
\langle 0\!\!\!\!\!&&\!\!\!\!\!|j_\mu^L(0)_{a_0c_0}|A,\theta_1,b_1,a_1;A,\theta_2,b_2,a_2;P,\theta_3,a_3,b_3;P,\theta_4,a_4,b_4\rangle\nonumber\\
&=&-\epsilon_{\mu\nu}(p_1+p_2+p_3+p_4)^\nu\frac{8\pi^2 i}{N}\nonumber\\
&&\times\left\{\frac{\tanh\left(\frac{\theta_{13}}{2}\right)}{(\theta_{14}+\pi i)(\theta_{23}+\pi i)(\theta_{24}+\pi i)}\right.\left(\delta_{a_0a_3}\delta_{a_1c_0}\delta_{a_2a_4}\delta_{b_1b_4}\delta_{b_2b_3}-\frac{1}{N}\delta_{a_0c_0}\delta_{a_1a_3}\delta_{a_2a_4}\delta_{b_1b_4}\delta_{b_2b_3}\right)\nonumber\\
&&+\frac{\tanh\left(\frac{\theta_{14}}{2}\right)}{(\theta_{13}+\pi i)(\theta_{23}+\pi i)(\theta_{24}+\pi i)}\left(\delta_{a_0a_4}\delta_{a_1c_0}\delta_{a_2a_3}\delta_{b_1b_3}\delta_{b_2b_4}-\frac{1}{N}\delta_{a_0c_0}\delta_{a_1a_4}\delta_{a_2a_3}\delta_{b_1b_3}\delta_{b_2b_4}\right)\nonumber\\
&&+\frac{\tanh\left(\frac{\theta_{23}}{2}\right)}{(\theta_{14}+\pi i)(\theta_{13}+\pi i)(\theta_{24}+\pi i)}\left(\delta_{a_0a_3}\delta_{a_1a_4}\delta_{a_2c_0}\delta_{b_1b_3}\delta_{b_2b_4}-\frac{1}{N}\delta_{a_0c_0}\delta_{a_2a_3}\delta_{a_1a_4}\delta_{b_1b_3}\delta_{b_2b_4}\right)\nonumber\\
&&+\frac{\tanh\left(\frac{\theta_{24}}{2}\right)}{(\theta_{14}+\pi i)(\theta_{13}+\pi i)(\theta_{23}+\pi i)}\left.\left(\delta_{a_0a_4}\delta_{a_1a_3}\delta_{a_2c_0}\delta_{b_1b_4}\delta_{b_2b_3}-\frac{1}{N}\delta_{a_0c_0}\delta_{a_2a_4}\delta_{a_1a_3}\delta_{b_1b_4}\delta_{b_2b_3}\right)\right\},\label{finalanswerfourparticle}
\eeq
where $\theta_{ij}=\theta_i-\theta_j$. The form factor with two incoming and two outgoing excitations can be found by using the S-matrix and crossing symmetry:
\beq
\langle P,\theta_2,b_2,a_2\!\!\!\!\!&;&\!\!\!\!\!A,\theta_4,a_4,b_4 \vert j_\mu^L(0)_{a_0c_0}\vert A,\theta_1,b_1,a_1;P,\theta_3,a_3,b_3\rangle\nonumber\\
&=&-\epsilon_{\mu\nu}(p_1+p_3-p_2-p_4)^\nu\frac{8\pi^2 i}{N}\nonumber\\
&&\times\left\{\frac{\tanh\left(\frac{\theta_{13}}{2}\right)}{(\theta_{14}+2\pi i)(\theta_{23}-2\pi i)(\theta_{24}+\pi i)}\right.\left(\delta_{a_0a_3}\delta_{a_1c_0}\delta_{a_2a_4}\delta_{b_1b_4}\delta_{b_2b_3}-\frac{1}{N}\delta_{a_0c_0}\delta_{a_1a_3}\delta_{a_2a_4}\delta_{b_1b_4}\delta_{b_2b_3}\right)\nonumber\\
&&+\frac{\coth\left(\frac{\theta_{14}}{2}\right)}{(\theta_{13}+\pi i)(\theta_{23}-2\pi i)(\theta_{24}+\pi i)}\left(\delta_{a_0a_4}\delta_{a_1c_0}\delta_{a_2a_3}\delta_{b_1b_3}\delta_{b_2b_4}-\frac{1}{N}\delta_{a_0c_0}\delta_{a_1a_4}\delta_{a_2a_3}\delta_{b_1b_3}\delta_{b_2b_4}\right)\nonumber\\
&&+\frac{\coth\left(\frac{\theta_{23}}{2}\right)}{(\theta_{14}+2\pi i)(\theta_{13}+\pi i)(\theta_{24}+\pi i)}\left(\delta_{a_0a_3}\delta_{a_1a_4}\delta_{a_2c_0}\delta_{b_1b_3}\delta_{b_2b_4}-\frac{1}{N}\delta_{a_0c_0}\delta_{a_2a_3}\delta_{a_1a_4}\delta_{b_1b_3}\delta_{b_2b_4}\right)\nonumber\\
&&+\frac{\tanh\left(\frac{\theta_{24}}{2}\right)}{(\theta_{14}+2\pi i)(\theta_{13}+\pi i)(\theta_{23}-2\pi i)}\left.\left(\delta_{a_0a_4}\delta_{a_1a_3}\delta_{a_2c_0}\delta_{b_1b_4}\delta_{b_2b_3}-\frac{1}{N}\delta_{a_0c_0}\delta_{a_2a_4}\delta_{a_1a_3}\delta_{b_1b_4}\delta_{b_2b_3}\right)\right\}\!\!.\label{fourparticlecrossed}
\eeq


\begin{thebibliography}{xx}

\bibitem{twoplusone} P. Orland, Phys. Rev. {\bf D 71} (2005) 054503. 
\bibitem{integrablePCSM} A.M. Polyakov and P.B. Wiegmann, Phys. Lett. {\bf 131 B} (1983) 121; P.B. Wiegmann, Phys. Lett. {\bf 141 B} (1984) 217; E. Abdalla, M.C.B. Ab- dalla and M. Lima-Santos, Phys. Lett. {\bf 140 B} (1984) 71; P.B. Wiegmann, Phys. Lett. {\bf 142 B} (1984) 173; L.D. Faddeev, N.Yu. Reshetikhin, Ann. Phys. {\bf 167} (1986) 227.
\bibitem{wiegmann}  P. B. Wiegmann, Phys. Lett. {\bf 142B}, 173 (1984).



\bibitem{horizontal} P. Orland, Phys. Rev. {\bf D74} (2006) 085001.
\bibitem{vertical} P. Orland, Phys. Rev. {\bf D77} (2008) 025035.

\bibitem{glueball} P. Orland, Phys. Rev. {\bf D 75} (2007) 101702.
\bibitem{multiparticle} A. Cortes Cubero, Phys. Rev. {\bf D 86} (2012) 025025. 
\bibitem{verlinde}H. Verlinde and E. Verlinde, Princeton University Preprint {\bf PUPT-1319}, hep-th/9302104 (1993).

\bibitem{MV}L. McLerran and R. Venugopalan, Phys. Rev. {\bf D49} (1994) 2233; {\bf D49} (1994) 3352; {\bf D50} (1994) 2225; {\bf D59} (1999) 094002.
\bibitem{miransky}V.A. Miransky and I.A. Shovkovy, Phys.Rev. {\bf D66} (2002) 045006.

\bibitem{nair}D. Karabali and V.P. Nair, Nucl. Phys. {\bf B464} (1996) 135; Phys. Lett. {\bf B379} (1996) 141; D. Karabali, C. Kim and V.P. Nair {\bf B524} (1998) 661; Phys. Lett. {\bf B434} (1998) 103; Nucl. Phys. {\bf B566} (2000) 331, Phys. Rev. {\bf D64} (2001) 025011.
\bibitem{greensite}J.P. Greensite, Nucl. Phys. {\bf B166} (1980) 113.
\bibitem{KNY}D. Karabali, V. P. Nair and A. Yelnikov, Nucl.Phys. {\bf B824} (2010) 387-414.

\bibitem{DMS} G. Delfino, G. Mussardo and P. Simonetti, Nucl. Phys. {\bf B 473} (1996) 469.
\bibitem{DM} G. Delfino and G. Mussardo, Nucl. Phys. {\bf B 516} (1998) 675.
\bibitem{DGM} G. Delfino, P. Grinza, and G. Mussardo, Nucl. Phys. {\bf B 737} (2006) 291.

\bibitem{konikadamov} R.M. Konik and Y. Adamov, Phys. Rev. Lett. {\bf 102} (2009) 097203.
\bibitem{konikjames}R.M. Konik and A.J.A. James Phys. Rev. {\bf B 87} (2013) 241103.
\bibitem{bhaseen} M.J. Bhaseen and A.M. Tsvelik, in {\bf From Fields to Strings; Circumnavigating Theoretical Physics}, Ian Kogan
memorial volumes, Vol. 1 (2004), pg. 661,
{\bf arXiv:cond-mat/0409602.}

\bibitem{kogut} J. Kogut and L. Susskind, Phys. Rev. D 11 (1975) 395. 
\bibitem{karowski} M. Karowski and P. Weisz, Nucl. Phys. {\bf B139}, 455 (1978).

\bibitem{zamolodchikov}A.B. Zamolodchikov and Al. B. Zamolodchikov, Nucl. Phys. B133 (1978) 525.
\bibitem{myselforland}A. Cortes Cubero and P. Orland, Phys. Rev. {\bf D 88} (2013) 025044.
\bibitem{oneplusone}A. Cortes Cubero and P. Orland, Phys. Rev. {\bf D 89} (2014) 085027.
\bibitem{mccoy} B.M. McCoy and T. T. Wu, Phys. Rev. {\bf D 18} (1978) 1259.
\bibitem{fonseca} P. Fonseca and A. B. Zamolodchikov, J. Stat. Phys, {\bf 110} (2003) 527.
\bibitem{renormalizedfield}P. Orland, Phys. Rev. {\bf D 84} (2011) 105005; Phys. Rev. {\bf D 86} (2012) 045023.
\bibitem{threeplusone} P. Orland, Phys. Rev. {\bf D77} (2008) 0560004.
\bibitem{TSA}[12] V. P. Yurov and Al. B. Zamolodchikov, Int. J. Mod. Phys
{\bf A 6}, 4557 (1991).
\bibitem{babujian} H. Babujian, A. Fring, M. Karowski, and A. Zapletal, Nucl. Phys. {B538}, 535 (1999); H. Babujian and M. Karowski, Nucl. Phys. {B620}, 407 (2002).
\bibitem{berg}B. Berg, M. Karowski and P. Weisz, Nucl. Phys. {\bf B134} (1978) 125.
\bibitem{kurak}V. Kurak and J.A. Swieca, Phys. Lett {\bf 82B} (1979) 289.

\bibitem{thesis} A. Cortes Cubero, PhD Thesis, Graduate School and University Center of the City University of New York, 2014.








\end{thebibliography}
\end{document}